\documentstyle[12pt,epsf,fleqn]{elsart}

\def \ni{\noindent}

\def \znbb {$0\nu\beta\beta$ }
\def \tnbb {$2\nu\beta\beta$ }
\def \be {\begin{equation}}
\def \ee {\end{equation}}
\def \HM {HEIDEL\-BERG-MOSCOW~}

\begin{document}
\small

\ni {\bf Corresponding author}\\
Prof. Dr. H.V. Klapdor-Kleingrothaus\\
Max-Planck-Institut f\"ur Kernphysik\\
Saupfercheckweg 1, D-69117 HEIDELBERG, GERMANY\\
Phone Office: +49-(0)6221-516-262, Fax: +49-(0)6221-516-540\\

\begin{frontmatter}

\title{Search for Neutrinoless Double Beta Decay with Enriched $^{76}{Ge}$ 
in Gran Sasso 1990-2003}

\author{H.V. Klapdor-Kleingrothaus}
\footnote{Spokesman of HEIDELBERG-MOSCOW (and GENIUS) Collaboration,\\
E-mail: H.Klapdor@mpi-hd.mpg.de,\\ Home-Page: $http://www.mpi-hd.mpg.de.non\_acc/$}, 
\protect\newline {I.V. Krivosheina  
\footnote{On leave of the Radiophysical-Research Institute, 
Nishnii-Novgorod, Russia}, A. Dietz,  and O. Chkvorets}

\address{Max-Planck-Institut f\"ur Kernphysik, PO 10 39 80,
  D-69029 Heidelberg, Germany}

\date{24.02.2004}

\begin{abstract}
	The results of the \HM experiment which searches 
	with 11\,kg of enriched $^{76}{Ge}$ 
	for double beta decay in the GRAN Sasso underground laboratory 
	are presented for the full running period August 1990 - May 2003. 
	The duty cycle of the experiment was $\sim$80\%, 
	the collected statistics is 71.7\,kg\,y. 
	The background achieved in the energy region 
	of the Q value for double beta decay 
	is 0.11\,events/\,kg\,y\,keV.
	The two-neutrino accompanied half-life is determined 
	on the basis of more than 100 000\,events. 
	The confidence level for the neutrinoless signal has been  
	{\bf improved to 4.2$\sigma$}.
\end{abstract}
\end{frontmatter}

\section{Introduction}
	Since 40 years large experimental efforts have gone 
	into the investigation of nuclear double beta decay 
	which probably is the most sensitive way to look for (total) 
	lepton number violation and probably the only way 
	to decide the Dirac or Majorana nature of the neutrino. 
	It has further perspectives 
	to probe also other types of beyond standard model physics. 
	This thorny way has been documented recently 
	in some detail 
\cite{KK60Y}.

	The half-lives to explore lying, with the order 
	of 10$^{25}$\,years, in a range  
	on 'half way' to that of proton decay, 
	the two main experimental problems were to achieve 
	a sufficient amount of double beta emitter material (source 
	strength) and to reduce the background in such experiment 
	to an extremely low level. 
	The final dream behind all these efforts was less 
	to see a standard-model allowed second-order effect 
	of the weak interaction in the nucleus - the 
	two-neutrino-accompanied decay mode -  
	which has been observed meanwhile for about ten nuclei - 
	but to see neutrinoless double beta decay, and with
	this a first hint of beyond standard model physics, 
	yielding at the same time a solution 
	of the absolute scale of the neutrino mass spectrum. 

	In this paper we describe the \HM experiment, 
	which was proposed in 1987 
\cite{Prop87-HM-HVKK},
	and which started operation in the Gran Sasso 
	Underground Laboratory with the first detector 
	in 1990.
	We report the result of the measurements 
	over the full period August 1990 until May 2003 
	(the experiment is still running, 
	at the time of submission of this paper). 
	The quality of the new data and of the present 
	analysis, which has been improved in various respects, 
	allowed us to significantly improve the investigation 
	of the neutrinoless double beta decay process, 
	and to deduce more stringent values on its parameters.	


	We are giving in a detailed paper 
\cite{New-Anal03}
	(for earlier papers see 
\cite{HDM97,HDM01})
	a full description of the experimental procedure, 
	the specific features and challenges 
	of the data acquisition and background reduction. 
	In this letter we mainly  
	present the data, taken with a total statistics of 71.7\,kg\,y 
	for the period August 2, 1990 - May 20, 2003, 
	and we present their analysis.

\section{Experimental Parameters}

	The \HM experiment 
	is, with five enriched (to (86\% $\div$ 88\%)) 
	high-purity p-type Germanium detectors, 
	of in total 10.96\,kg of active volume,  
	using the largest source strength of all double beta 
	experiments at present, and has reached a record
        low level of background. 
	It is since ten years now the most sensitive
	double beta decay experiment worldwide. 
	The detectors were the first {\it high-purity} Ge detectors 
	ever produced.	
	The degree of enrichment has been checked by investigation 
	of tiny pieces of Ge {\it after} crystal production using 
	the Heidelberg MP-Tandem accelerator as a mass spectrometer. 
	Since 2001 the experiment has been operated only 
	by the Heidelberg group, 
	which also performed the analysis of the experiment 
	from its very beginning.

	We start by listing up some of the most essential 
	features of the experiment.

\begin{enumerate}
\item
	Since the sensitivity for the \znbb half-life is
	$T^{0\nu}_{1/2} \sim a  \times \epsilon \sqrt{\frac{Mt}{\Delta EB}}$ 
	 (and 
	$\frac{1}{\sqrt{T^{0\nu}}} \sim \langle m_\nu \rangle$), 
	with $a$ denoting the degree of enrichment, $\epsilon$ 
	the efficiency of the detector for detection of a double beta event, 
	$M$ the detector (source) mass, $\Delta E$ the energy resolution, 
	$B$ the background and $t$ the measuring time, 
	the sensitivity of our 11\,kg {\it of enriched} $^{76}{Ge}$ 
	experiment corresponds to that of an at least 1.2\,ton 
	{\it natural} Ge experiment. After enrichment -  
	the other most important parameters of a $\beta\beta$ 
	experiment are:	energy resolution, 
	background and source strength.

\item
	The high energy resolution of the Ge detectors of 0.2$\%$ 
	or better, assures that there 
	is no background for a \znbb line from the two-neutrino 
	double beta decay in this experiment 
	(5.5 $\times$ 10$^{-9}$\,events expected 
	in the energy range 2035$\div$2039.1\,keV), in contrast to most 
	other present experimental approaches, 
	where limited energy resolution is a severe drawback.

\item
	The efficiency of Ge detectors for detection 
	of \znbb decay events is close to 100\,$\%$ (95\%, see 
\cite{Diss-Dipl}(a)).

\item
	The source strength in this experiment of 11\,kg 
	is the largest source strength ever operated in 
	a double beta decay experiment.

\item
	The background reached in this experiment, is  
	0.113$\pm$0.007\,events/kg\,y\,keV (in the period 1995-2003)  
	in the \znbb decay region 
	(around Q$_{\beta\beta}$). 
	This is the lowest limit ever obtained 
	in such type of experiment.

\item
	The statistics collected in this experiment during 
	13 years of stable running is the largest ever collected 
	in a double beta decay experiment. 
	The experiment took data during $\sim$ 80\% of its installation time.

\item
	The Q value for neutrinoless double beta decay has been 
	determined recently with  
	high precision 
\cite{New-Q-2001,Old-Q-val}.

\end{enumerate}

\section{Background of the experiment}

	The background of the experiment consists of

\begin{enumerate}
\item 
	primordial activities of the natural 
	decay chains from $^{238}{U}$, $^{232}{Th}$, and $^{40}{K}$;  

\item 	
	anthropogenic radio nuclides, like $^{137}{Cs}$, 
	$^{134}{Cs}$, $^{125}{Sb}$, $^{207}{Bi}$;

\item 
	cosmogenic isotopes, produced by activation 
	due to cosmic rays during production and transport. 

\item 	
	the bremsstrahlungs
	spectrum of $^{210}{Bi}$ (daughter of $^{210}{Pb}$); 

\item 	
	elastic and inelastic neutron scattering, and 

\item 
	direct muon-induced events.
\end{enumerate}

	The detectors, except detector No. 4, 
	are operated 
	in a common Pb shielding of 30 cm, which consists of 
	an inner shielding of 10 cm radiopure LC2-grade Pb followed 
	by 20 cm of Boliden Pb. The whole setup is placed 
	in an air-tight steel box 
	and flushed 
	with radiopure nitrogen in order to suppress the $^{222}{Rn}$ 
	contamination of the air. 
	The shielding has been improved in the course of the measurement. 
	The steel box is since 1994 centered 
	inside a 10-cm boron-loaded polyethylene shielding 
	to decrease the neutron flux from outside. 
	An active anticoincidence shielding is placed on the top 
	of the setup since 1995 
	to reduce the effect of muons.
	Detector No. 4 is installed in a separate setup, which has 
	an inner shielding of 27.5 cm electrolytical Cu, 20 cm lead, 
	and boron-loaded polyethylene shielding below the steel box,  
	but no muon shielding.

	The setup has been kept hermetically closed since 
	installation 
	of detector 5 in February 1995. Since then 
	no radioactive 
	contaminations of the inner of the experimental setup by air 
	and dust from the tunnel could occur.

	A detailed description of the remaining background, 
	and corresponding Monte Carlo simulations with GEANT4 are given in 
\cite{KK-Doer03}.
	Important result is, that 
	there are no background $\gamma$-lines 
	to be expected at the position of an expected \znbb line, 
	according to this Monte Carlo analysis 
	of radioactive impurities in the experimental setup 
\cite{KK-Doer03}
	and according to the compilations in 
\cite{Tabl-Isot96}.
	Some $\gamma$--peaks in the simulated spectrum 
	near Q$_{\beta\beta}$ 
	are emitted from the isotope $^{214}$Bi. 
	So from $^{214}{Bi}$ 
	($^{238}U$-decay chain) 
	in the vicinity of the Q-value of the double beta decay of
	Q$_{\beta\beta}$~=~2039\,keV, 
	according to the Table of Isotopes 
\cite{Tabl-Isot96},
	weak lines should be expected at 
	2010.7, 2016.7, 2021.8 and 2052.9\,keV (see 
\cite{Bi-KK03-NIM}). 
	It should be especially noted, that also neutron 
	capture reactions $^{74}{Ge}(n,\gamma)^{75}{Ge}$ and 
	$^{76}{Ge}(n,\gamma)^{77}{Ge}$ 
	and subsequent radioactive decay have been simulated by GEANT4 in 
\cite{KK-Doer03}. 
	The simulation yields, in the range 
	of the spectrum 
	1990 - 2110\,keV, for the total contribution 
	from decay of $^{77}{Ge}$ 0.15\,counts. 
	So in particular, the 2037.8\,keV transition 
	in the $\gamma$-decay following ${\beta}^{-}$ decay 
	of  $^{77}{Ge}$ 
\cite{Tabl-Isot96}
	is not expected to be seen in the measured spectra. 
	It would always be accompanied by a 10\,times stronger 
	line at 2000.2\,keV 
\cite{Tabl-Isot96}, 
	which is not seen in the spectrum (Figs. 
\ref{fig:Low-HightAll90-03},\ref{fig:Sum90-95-03-Scan}).

\section{Control of stability of the experiment and data taking}

	To control the stability of the experiment, a calibration 
	with a $^{228}{Th}$ and 
\protect\newline a $^{152}{Eu}$+$^{228}{Th}$ source, 
	has been done weekly.
	High voltage of the detectors, temperature in the detector 
	cave and the computer room, the nitrogen flow flushing  
	the detector boxes to remove radon, the muon anticoincidence signal,  
	leakage current of the detectors, overall and individual 
	trigger rates are monitored daily.  
	The energy spectrum is taken in parallel in 8192 channels in the  
	range from threshold up to about 3 MeV, and in 
	a spectrum up to about 8 MeV.  
	Data taking since November 1995 is done by a CAMAC system, 
	and CETIA processor in event by event mode. 
	To read out the used 
	250\,MHz flash ADCs 
	of type Analog Devices 9038 JE (in DL515 modules), 
	which allow digital measurement of pulse shapes 
	for the four largest detectors (for later off-line analysis), 
	a data acquisition system on VME basis has been developed 
\cite{Diss-Dipl}(c). 
	The resolution of the FADC's is 8\,bit, 
	and thus not sufficient for a measurement of energies. 
	The energy signals for high and low-energy spectra 
	are recorded with 13\,bit ADC's developed at MPI Heidelberg. 
	The acquisition system 
	used {\it until} 1995 (VAX/VMS) did not allow to measure 
	the pulse shapes of all detectors because 
	of the too low data transfer rates.
	For the period 1990-1995 we prepared 
	the final sum spectrum taken with the former electronics shortly 
	after this measuring period, setting the proper conditions 
	for data acceptance 
\cite{Diss-Dipl}(a,b,c).
	Since 1995, in total we have taken with the since 
	then new electronics, 2142 runs (10 513 data 
	sets for the five detectors) 
	({\it without} calibration measurements), 
	the average length of which was about one day.
	From these raw data runs and data sets 
	only those are considered 
	for further analysis which fulfill the following conditions:

\begin{enumerate}
\item
	no coincidence with another Ge detector;

\item
	no coincidence with plastic scintillator (muon shield);

\item
	no deviation from average count rate of each detector 
	more \protect\newline than $\pm 5\sigma$;

\item
	only pulses with ratios of the energy determined 
	in the spectroscopy branch, and the area under 
	the pulse detected in the {\it timing} branch 
	(EoI values 
\cite{KKMaj99}), 
	within a $\pm$3$\sigma$ range around the mean value 
	for each detector are accepted;

\item
	we ignore for each detector the first 200\,days of operation, 
	corresponding to about three half-lifes 
	of $^{56}{Co}$ ~~($T_{1/2}^{0\nu}$=77.27\,days), 
	to allow for decay of short-lived radioactive impurities;

\item
	ADC's or other electronics units were working properly 
	(corrupted data are excluded).

\end{enumerate}

	From the totally registered 951 044\,events 
	in the spectrum measured since 1995, 
	there remain 
	finally 786 652\,events (9570 data sets).
	From these we find in the range 2000 to 2060\,keV 
	around Q$_{\beta\beta}$ -- 562~\,events.

\section{$Q$ Value}

	The expectation for a \znbb signal would be a sharp line 
	at the $Q$ value of the process.
	Earlier measurements gave $Q_{\beta\beta}$=(2040.71 $\pm$ 0.52)\,keV, 
	(2038.56 $\pm$ 0.32)\,keV and (2038.668 $\pm$ 2.142)\,keV
\cite{Old-Q-val}. 
	The recent precision measurement of 
\cite{New-Q-2001}
	gave a value of (2039.006 $\pm$ 0.050)\,keV.

\section{Development of the experiment}

	Table 
\ref{activities} 
	shows the collected statistics of the experimental setup and 
	of the background number in the different data acquisition 
	periods for the five enriched detectors of the experiment 
	for the total measuring time August 1990 to May 2003.


\begin{table}[th]
\caption{\label{activities}Collected statistics and 
	background numbers 
	in the different data acquisition periods for the enriched detectors
	of the HEIDELBERG-MOSCOW experiment for the period 1990 - 2003. 
	${^*)}$ background determined by averaging counting 
	rate in interval 2000-2100\,keV (counts/keV y kg); 
	background determined from 11.1995 till 05.2003
	at Q$_{\beta\beta}$ from {\it fit}  
	of spectrum in range 2000-2060\,keV 
	is (0.113$\pm$0.007)\,events/\,keV y kg (all numbers without PSA).}

\vspace{0.2cm}
\begin{center}
\renewcommand{\arraystretch}{1.5}
\setlength\tabcolsep{1.pt}
\begin{tabular}{|c|c|c|c|c|c|c|c|c|}
\hline
	Detec-		&	
Life Time		&	Date		&	
Backg- 	&	PSA	&	
Life Time		&	Date		&	
Backg- 	&	PSA	\\
	tor		&		
accepted 	&	Start End	&	
round${^*)}$ 	&	&	
accepted 	&	Start End	&	
round${^*)}$ 	&	\\
		No.	&	(days)
 		&	&			
&		&	(days)	&	&			&	\\
\hline
\hline
	1	&	930.9	&	8/90-8/95	&
	0.31	&	no	
&	2090.61	&	11/95-5/03	&
	0.20	&	no	\\
	2	&	997.2	& 9/91-8/95	&
	0.21	&	no	
&	1894.11	& 11/95-5/03	&
	0.11	&	yes	\\
	3	&	753.1	& 9/92-8/95	&
	0.20	&	no	
&	2079.46	& 11/95-5/03	&
	0.17	&	yes	\\
	4	&	61.0	& 1/95-8/95	&
	0.43	&	no	
&	1384.69	& 11/95-5/03	&
	0.21	&	yes	\\
	5	&	-	& 12/94-8/95	&
	0.23	&	no	
&	2076.34	& 11/95-5/03	&
	0.17	&	yes	\\
\hline
	&	\multicolumn{4}{|c|}
{\bf Over period 1990 - 1995}	&
\multicolumn{4}{|c|}
{\bf Over period 1995-2003}\\
	&	\multicolumn{4}{|c|}
{Accepted life time = {\bf 15.05\,kg\,y}} 	&
\multicolumn{4}{|c|}{Accepted life time = {\bf 56.655\,kg\,y}} 
\\
\hline
\end{tabular}
\end{center}
\end{table}


\vspace{0.5cm}
\section{Reliability of data acquisition and data}

\begin{figure}[ht]

\epsfysize=55mm\centerline{\epsffile{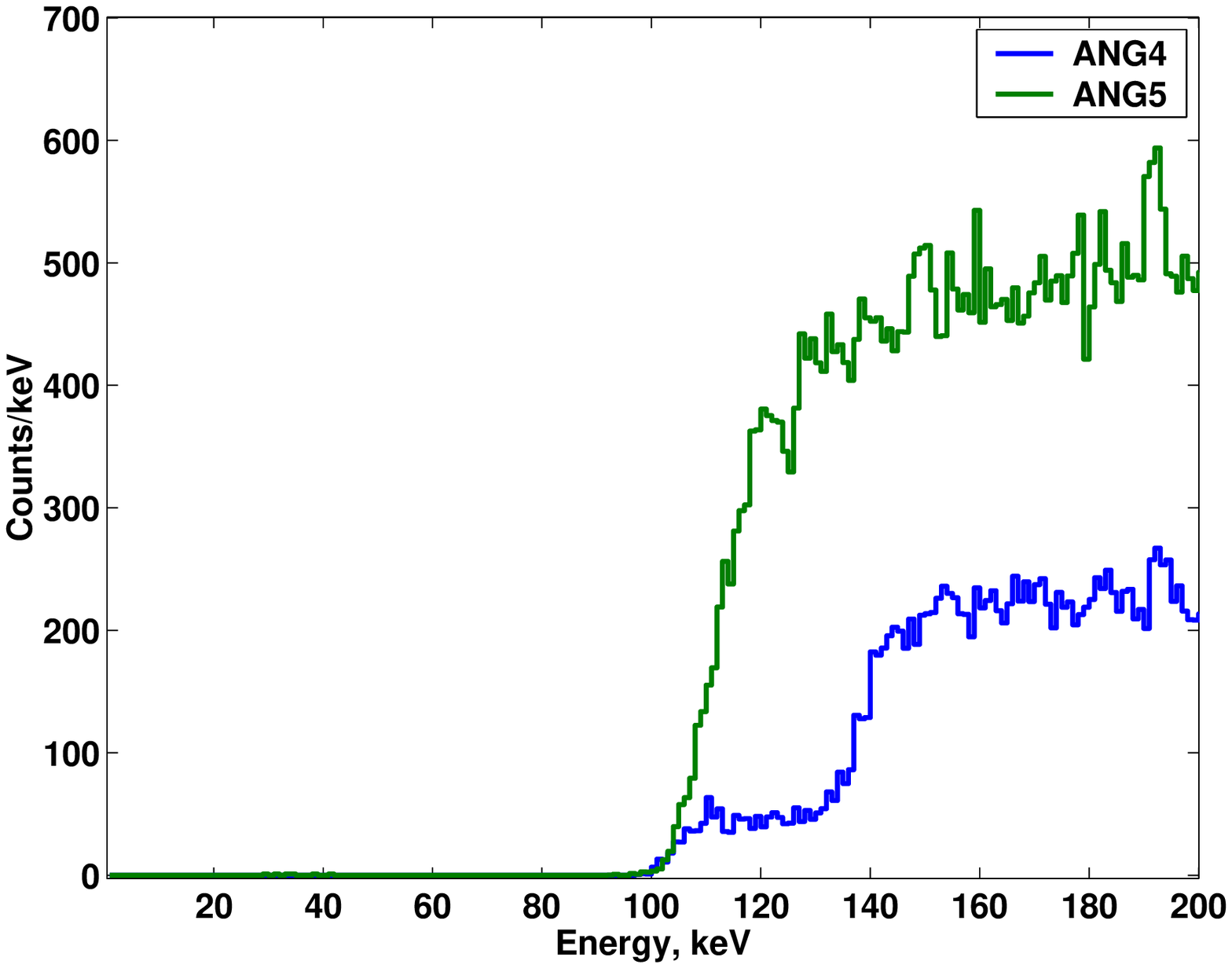}\hspace{1.cm}\epsfysize=55mm\epsffile{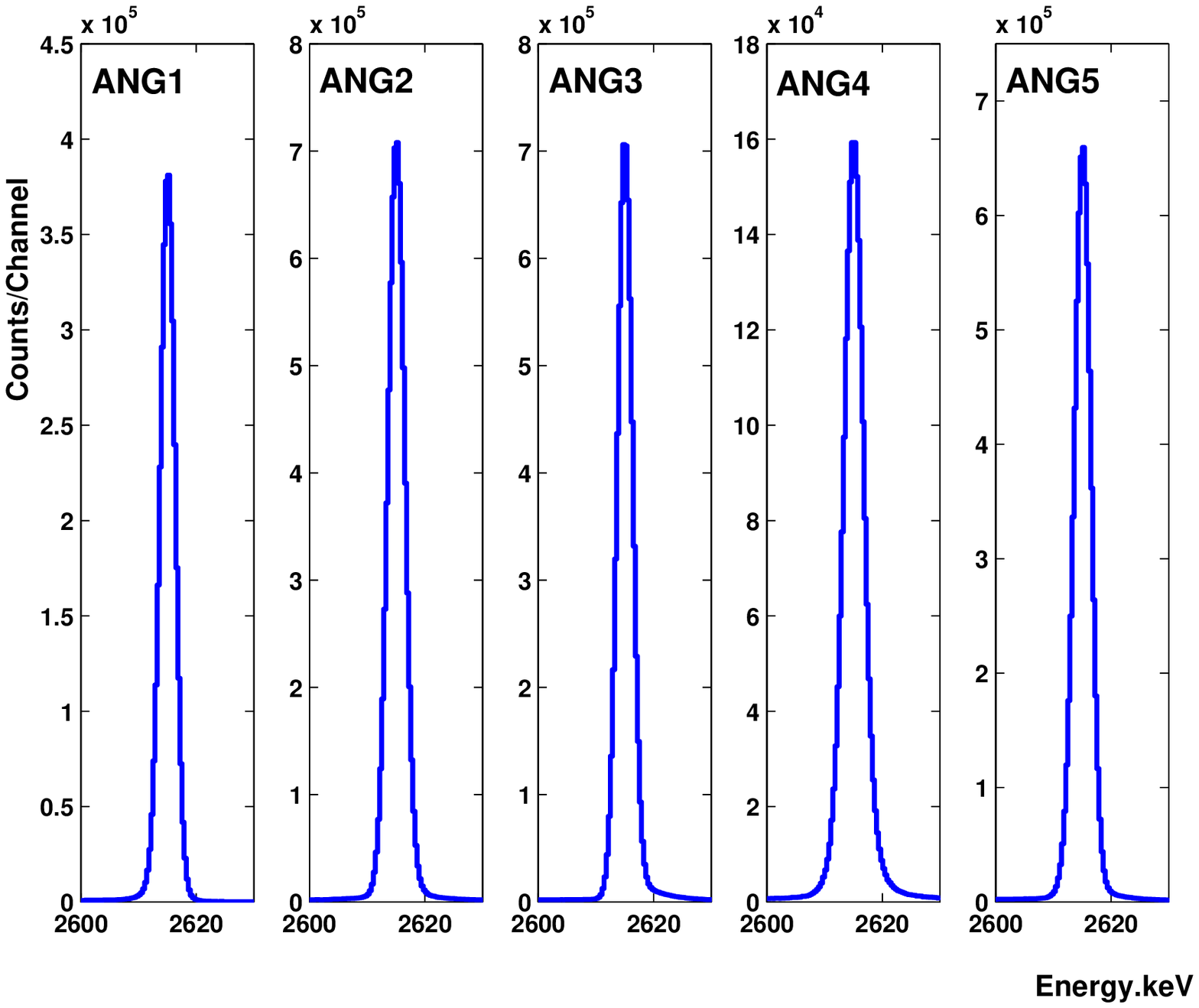}}
\caption[]{
	Left: Threshold ranges for detectors 4 and 5 
	in the spectrum measured during 1995 - 2003. 
	Right: {\it Sum} of the ($\sim$360) weekly calibration spectra 
	for the 2614.5\,keV line from  
	$^{228}{Th}$ over the measuring time 1995 - 2003.}
\label{fig:Thres-Range5det}
\end{figure}

	We have carefully checked all data before starting the analysis. 
	For details we refer to 
\cite{New-Anal03}. 
	We checked the proper operation of the thresholds 
	in all ADC's.
	Fig. 
\ref{fig:Thres-Range5det} 
	shows the threshold ranges for detectors 4 and 5, 
	from the list mode data acquisition 
	of the measurement, for the data used in the analysis.   
	The effect discussed in 
\cite{Kurch03} 
	(their Fig. 4) is not present in our analysis. 
	Also the other effects discussed in 
\cite{Kurch03}
	are not present in the analysis.
	A detailed analysis shows that these effects arise 
	from including corrupt data into the analysis (see also 
\cite{New-Anal03}).

\begin{figure}[ht]
\epsfysize=50mm\centerline{\epsffile{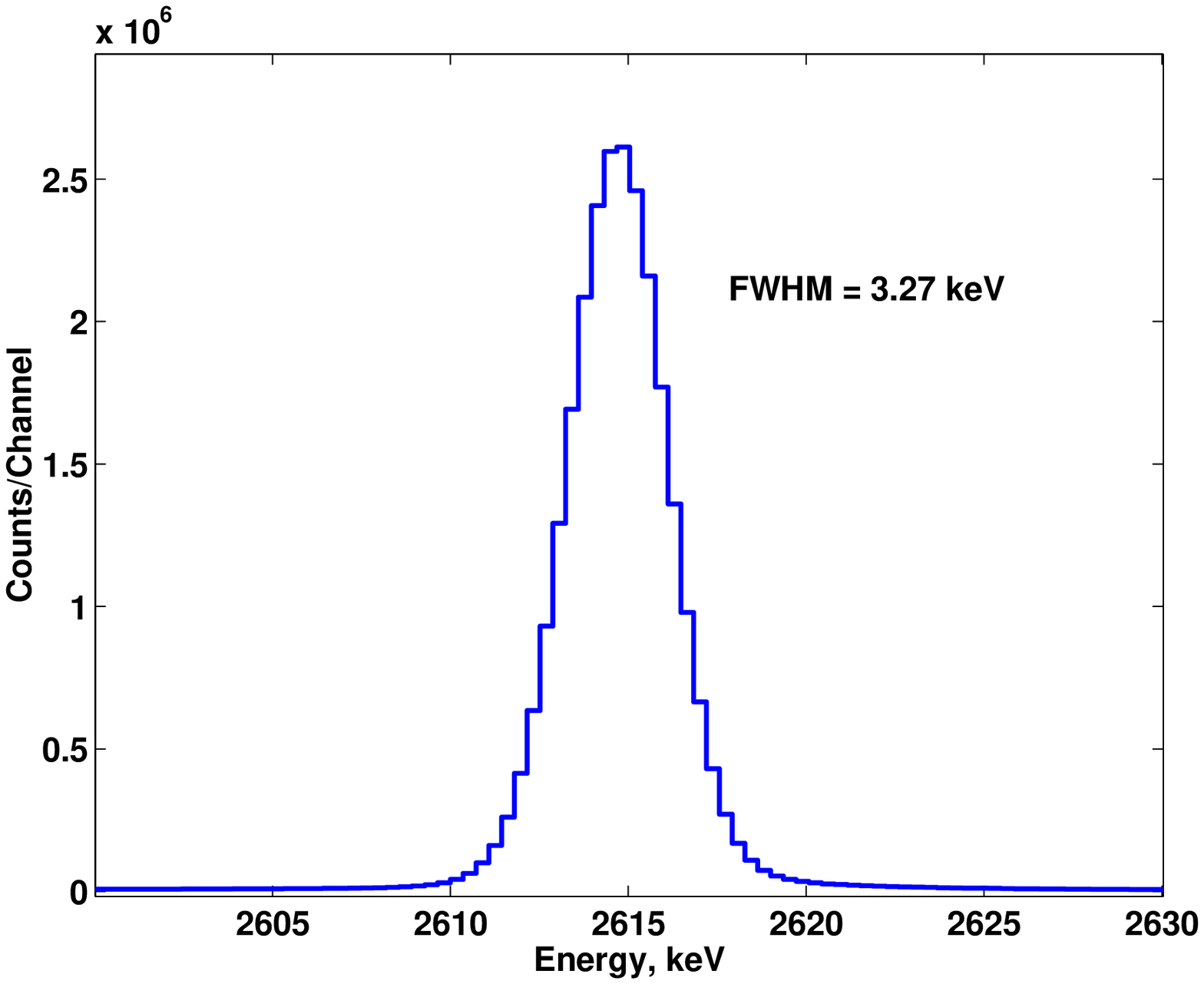}\hspace{1.cm}
\epsfysize=52mm\epsffile{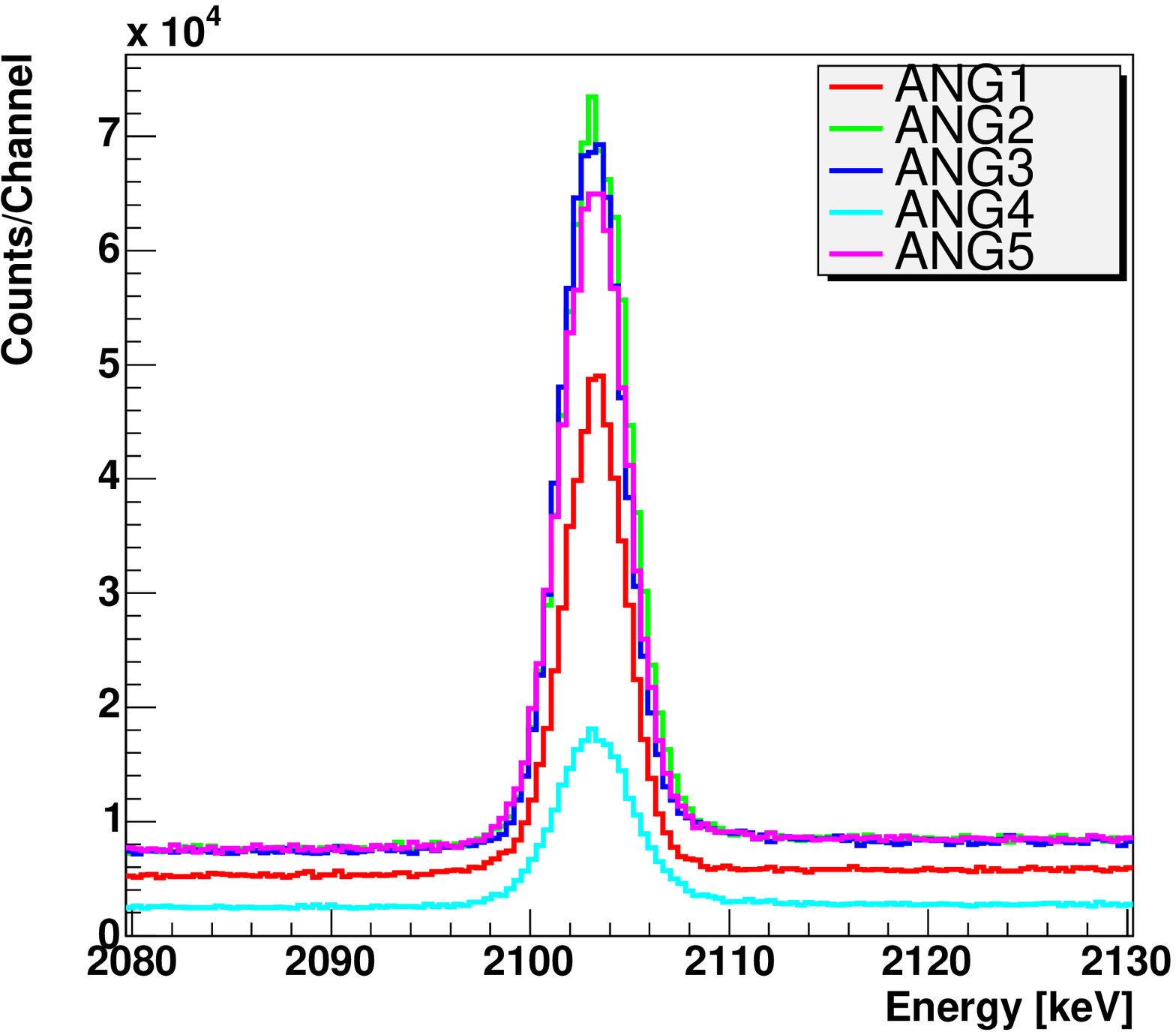}}
\caption[]{Left: Sum of the sum of the weekly calibration spectra 
	of all five detectors shown in Fig. 
\ref{fig:Thres-Range5det}(right) ($\sim$1800 spectra). 
	Right: Sum of calibration spectra for each detector 
	(absolute intensities) for the 2103.5\,keV line 
	for the period 1995 - 2003  
	(from top to bottom detectors: 2,3,5,1,4).
	}
\label{fig:Bi-2614-5det}
\end{figure}

        The stability 
        of the experiment over the years and of the energy resolution 
        and calibration is demonstrated in the following figures 
	for the two $^{228}{Th}$ lines used for calibration 
	in the range of Q$_{\beta\beta}$:  
	Fig. 
\ref{fig:Thres-Range5det} (right) 
	shows the {\it sum} of all weekly calibration spectra 
	($\sim${\it 360 spectra} for {\it each} detector) 
	for the 2614.5 keV Th line. 
	Fig. 
\ref{fig:Bi-2614-5det}(left)
	shows
        the resolution and energy position after {\it summing} 
	the spectra {\it of the five detectors} 
	($\sim$ {\it 1800 spectra}). 
	The integrated resolution over 8\,years of measurement 
	is found to  be ~ 3.27 \,keV, i.e. better than 
	in our earlier analysis of the data until May 2000 
\cite{KK02,KK02-Found-PN}.  
	This is a consequence also of the refined summing procedure 
	we used for the individual 9 570 data sets.

	Fig. 
\ref{fig:Bi-2614-5det} (right)  
	shows as another example the {\it sum} 
	of all weekly calibration spectra for 
        the 2103.5\,keV line for each detector. 
	The agreement of the energy positions 
	for the different detectors is seen to be very good.

	We also checked the arrival time distribution of the events, 
	and proved that this is not affected by 
	any technical operation during the measurement, 
	such as the calibration procedure 
	(introducing the Th source into the detector chamber 
	through a thin tube) and simultaneously 
	refilling of liquid nitrogen 
	to the detectors etc. 
	No hint is seen for introduction of any radioactive 
	material from the exterior. 
	The measured distribution corresponds 
	to a uniform distribution in time on a 99\% c.l. 
\cite{New-Anal03}.
	This has been proven by a Kolmogorov-Smirnov goodness-of-fit 
	test.


\section{Data and Analysis}

\begin{figure}[t]
\epsfysize=91mm\centerline{\epsffile{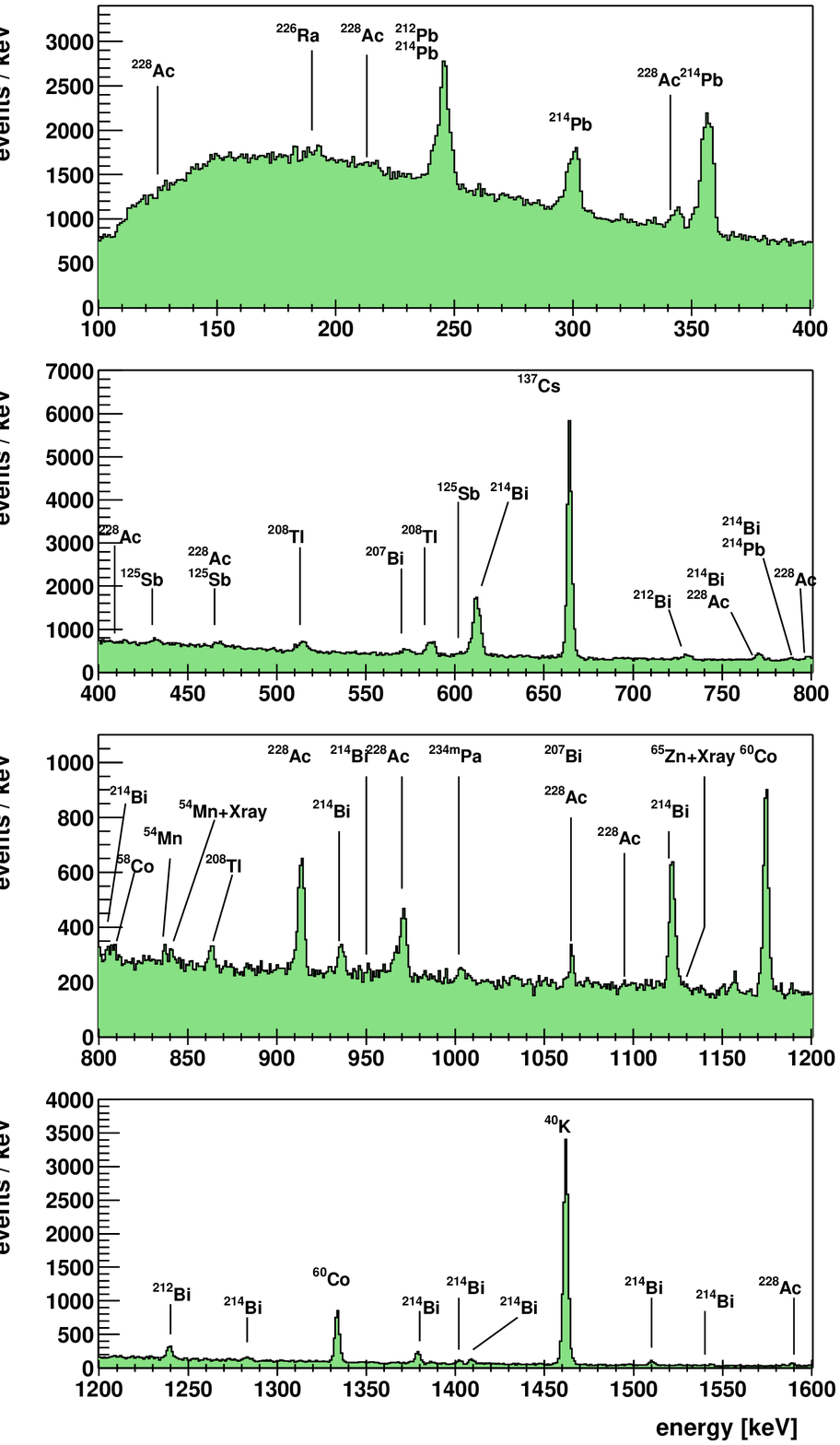}\hspace{1.cm}\epsfysize=90mm\epsffile{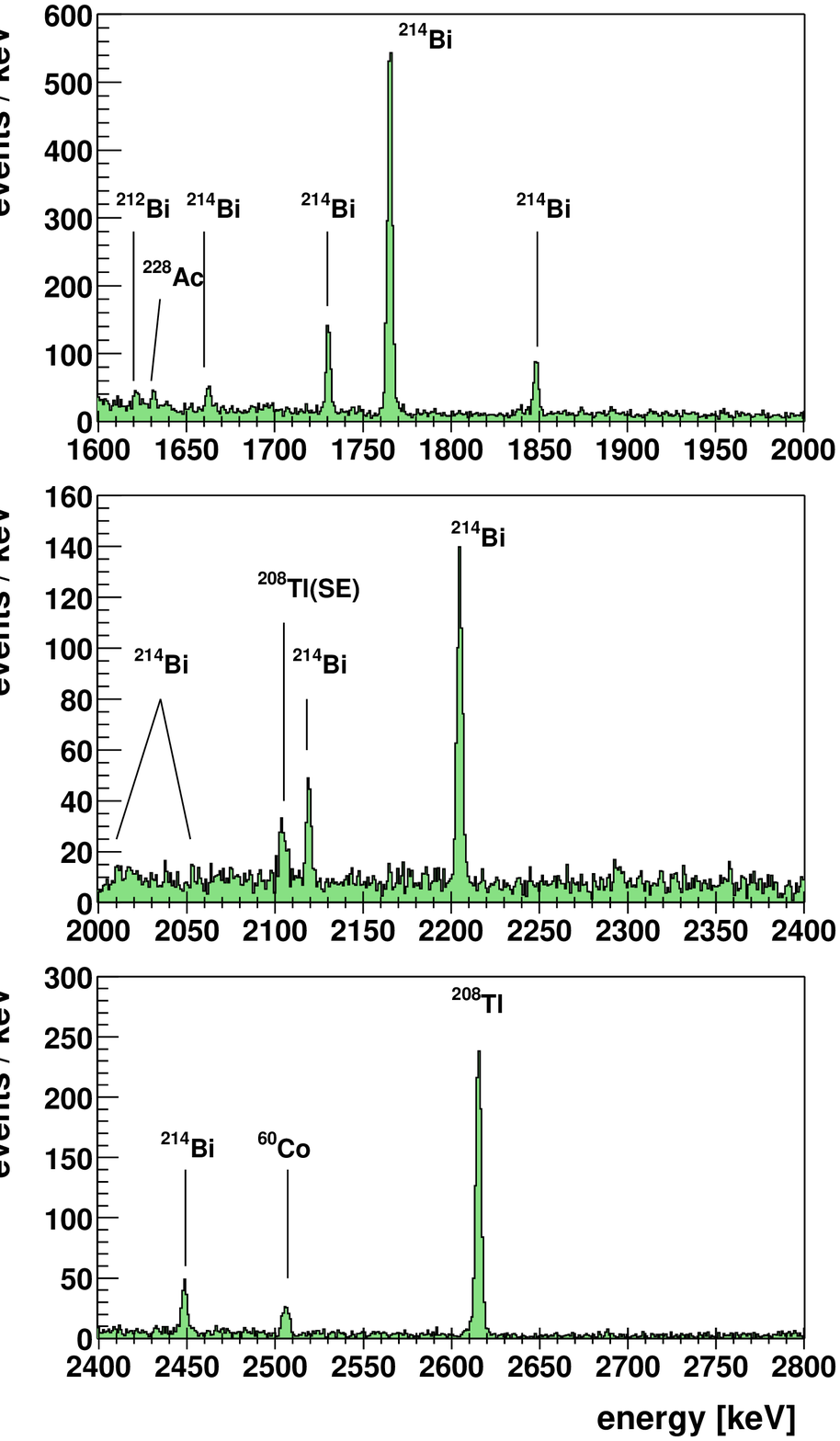}}
\caption[]{The total sum spectrum measured 
	over the full energy range 
	with all five detectors (in total 10.96\,kg enriched 
	in $^{76}{Ge}$  to 86\%) - for the period 
	August 1990 to May 2003. 
}
\label{fig:Low-HightAll90-03}
\end{figure}

	Fig. 
\ref{fig:Low-HightAll90-03} 
	shows the total sum spectrum measured over the full energy range 
	of all five detectors for the period 
	August 1990 to May 2003. 
	All identified lines are indicated with their source 
	of origin (for details see 
\cite{KK-Doer03},\cite{Diss-Dipl}(e,f)).

\begin{figure}[ht]
\epsfysize=43mm\centerline{\epsffile{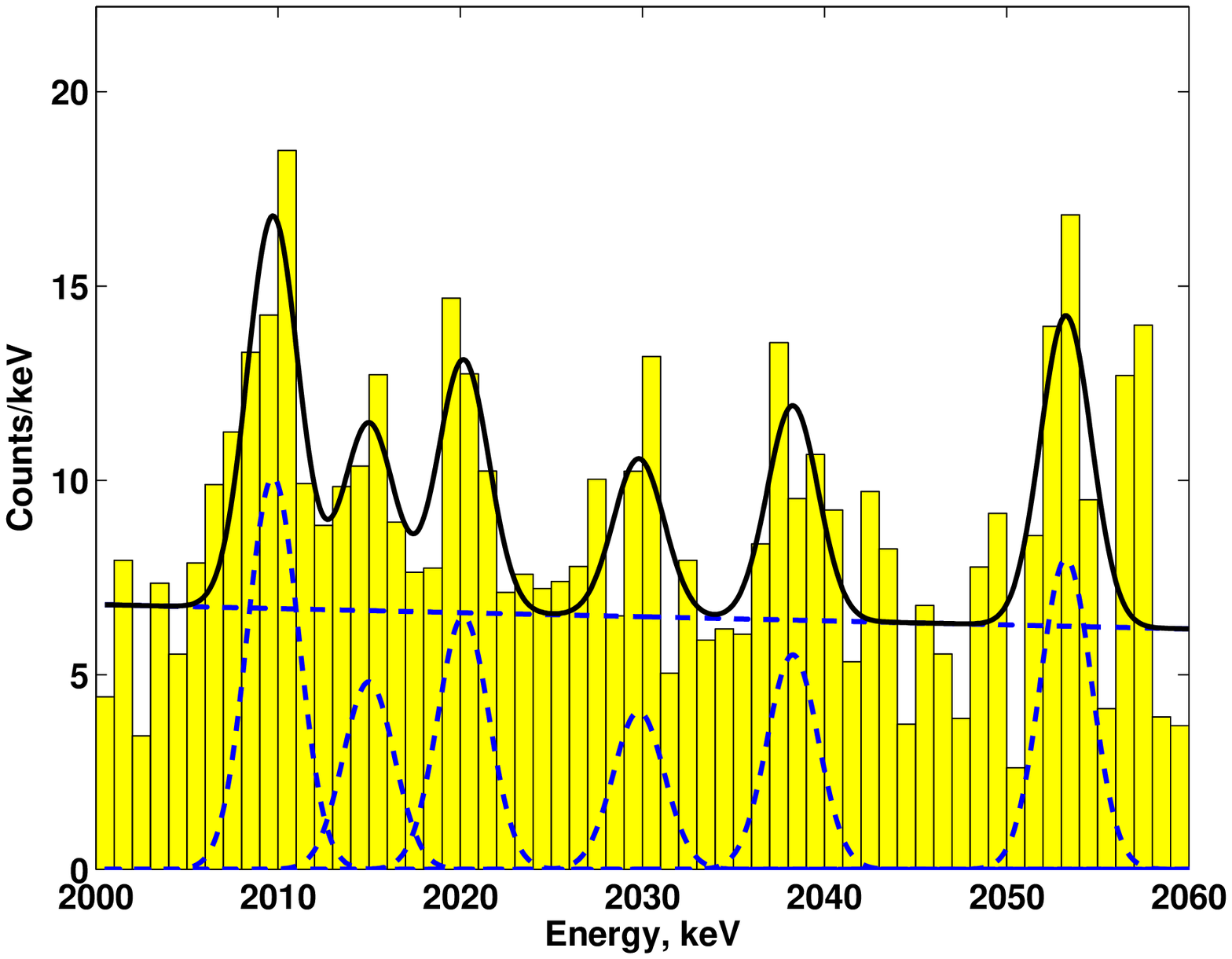}\hspace{0.5cm}\epsfysize=43mm\epsffile{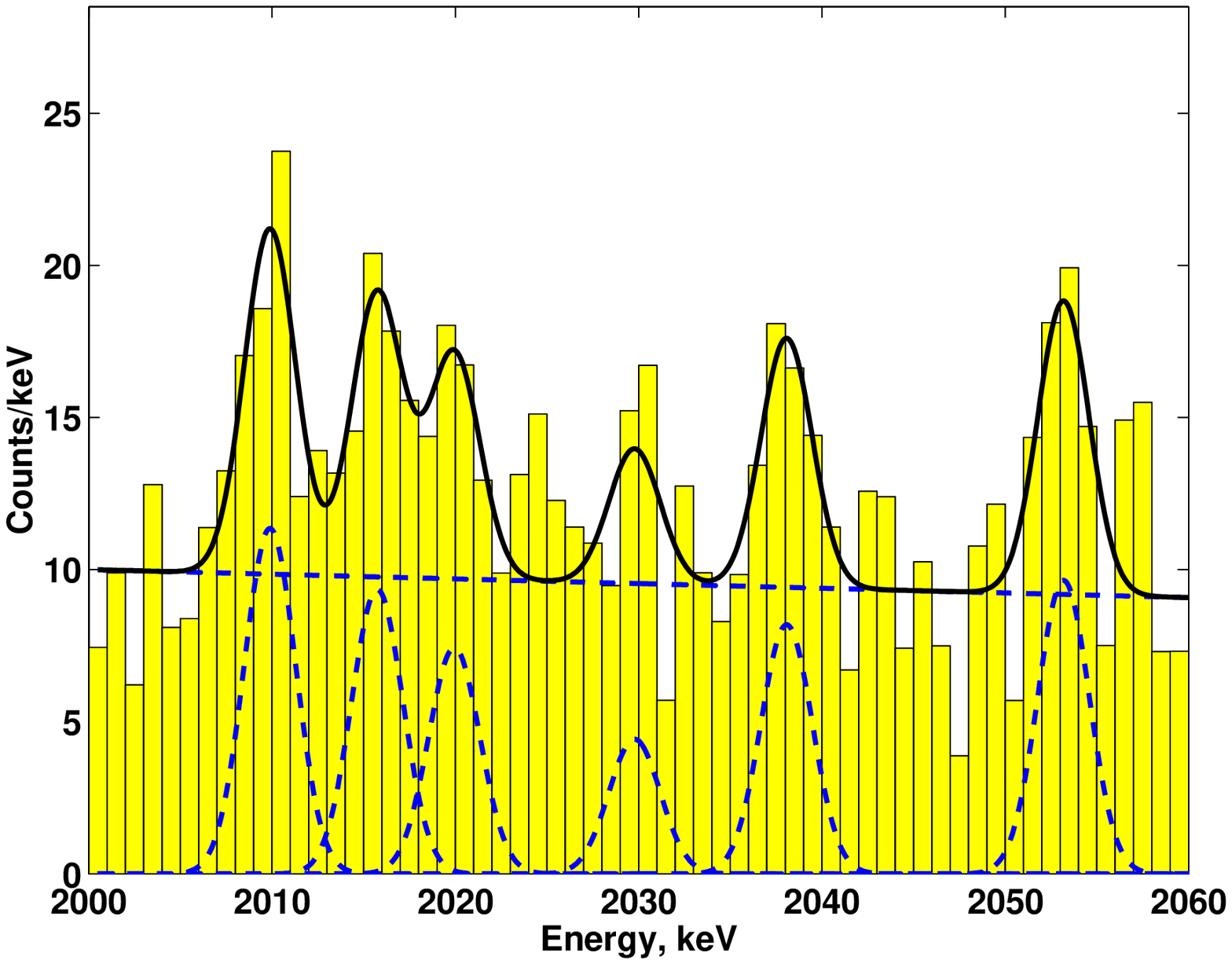}}
\epsfysize=43mm\centerline{\epsffile{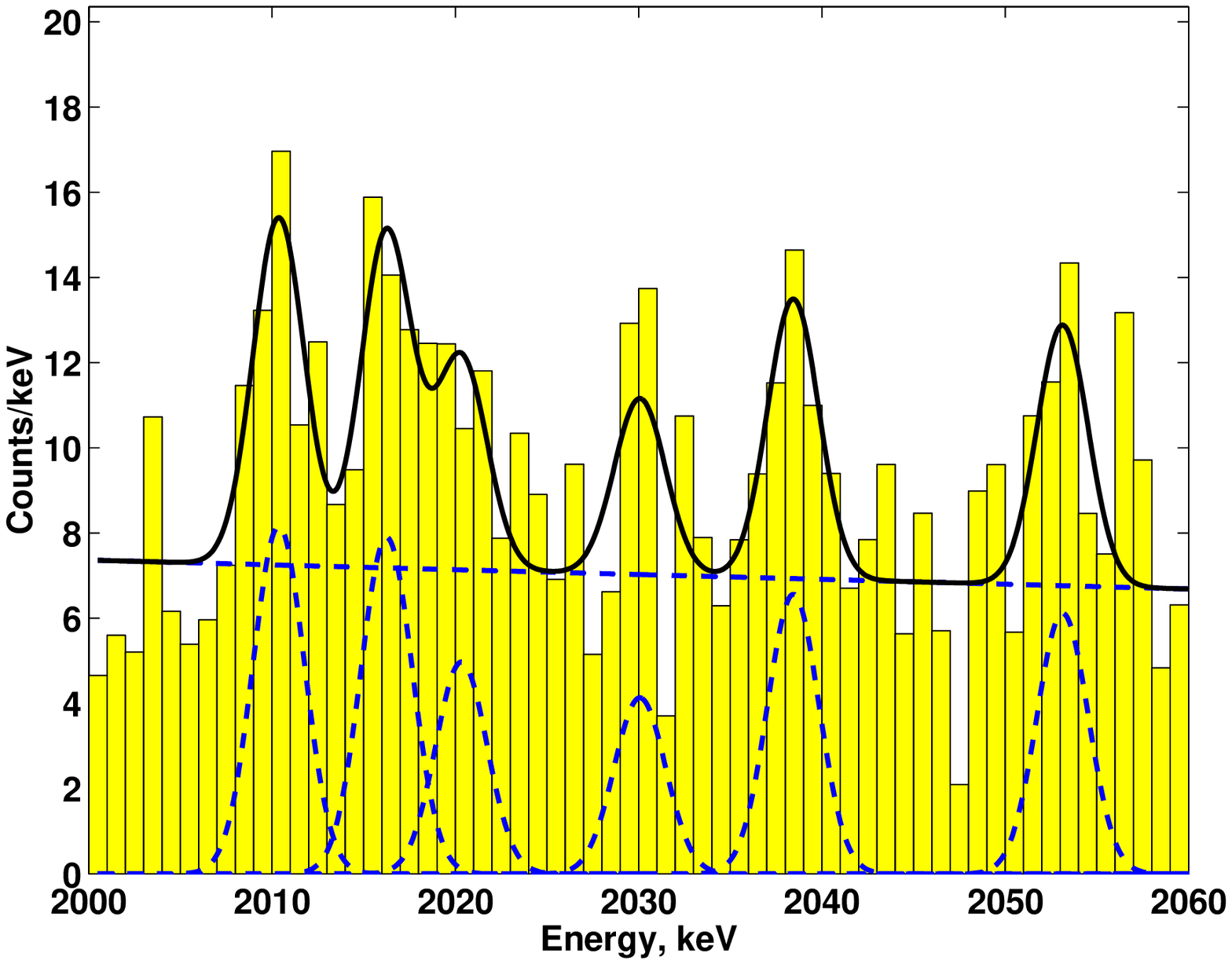}\hspace{0.7cm}\epsfysize=49mm\epsffile{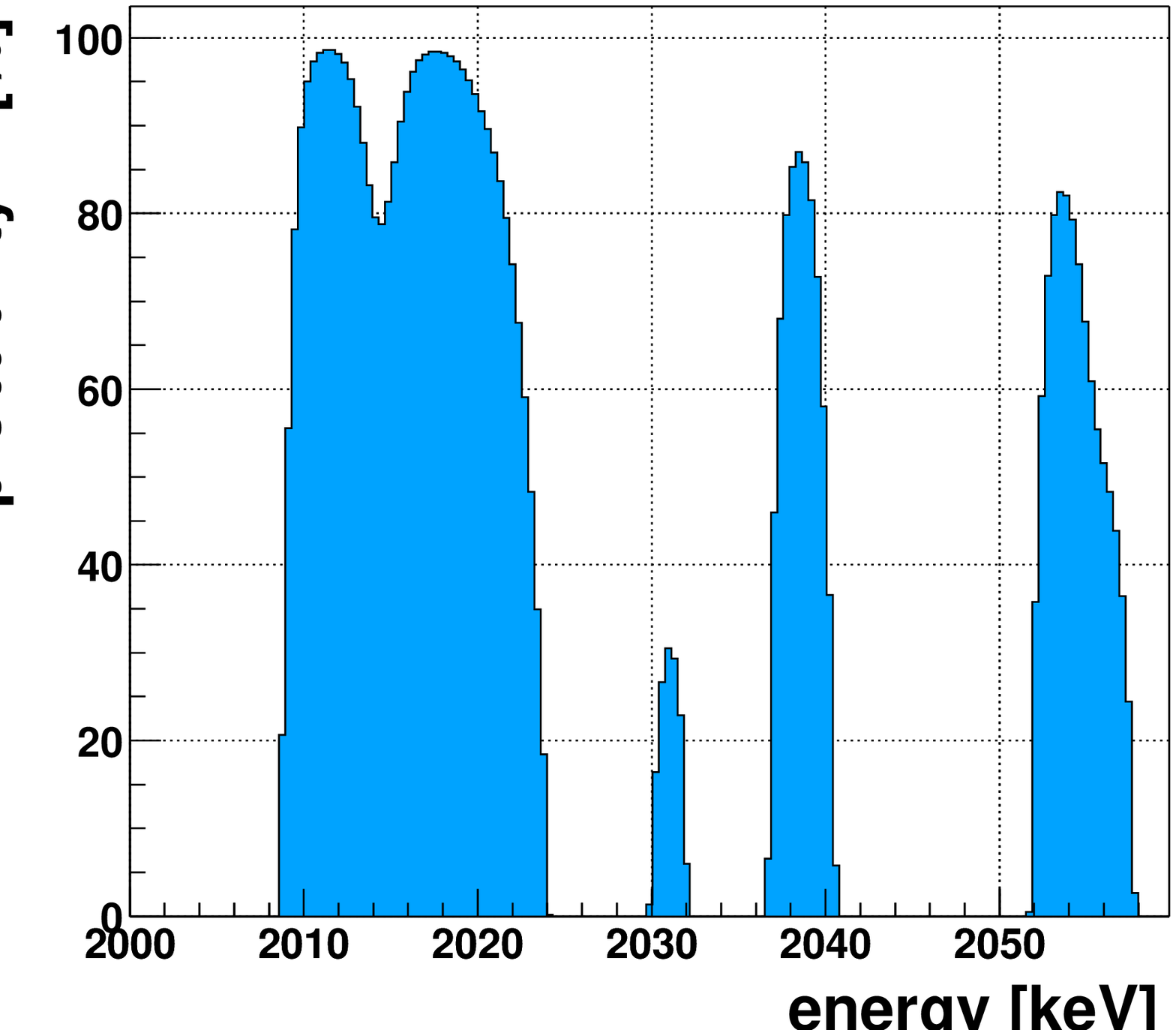}}
\caption[]{
    	The total sum spectrum of all five detectors 
	(in total 10.96\,kg enriched in $^{76}{Ge}$), 
	in the range 2000 - 2060\,keV 
	and its fit, for the periods: 
	Top: left - August 1990 to May 2000 (50.57\,kg\,y);  
	right - August 1990 to May 2003 (71.7\,kg\,y). 
	Bottom: left - November 1995 to May 2003 (56.66\,kg\,y);  
	right - scan for lines in the spectrum shown on the left, 
	with the MLM method (see text). 
	The Bi lines at 2010.7, 2016.7, 2021.8 and 2052.9\,keV 
	are seen, and in addition a signal at $\sim$ 2039\,keV.
}
\label{fig:Sum90-95-03-Scan}
\end{figure}

\begin{figure}[ht]
\epsfysize=40mm\centerline{\epsffile{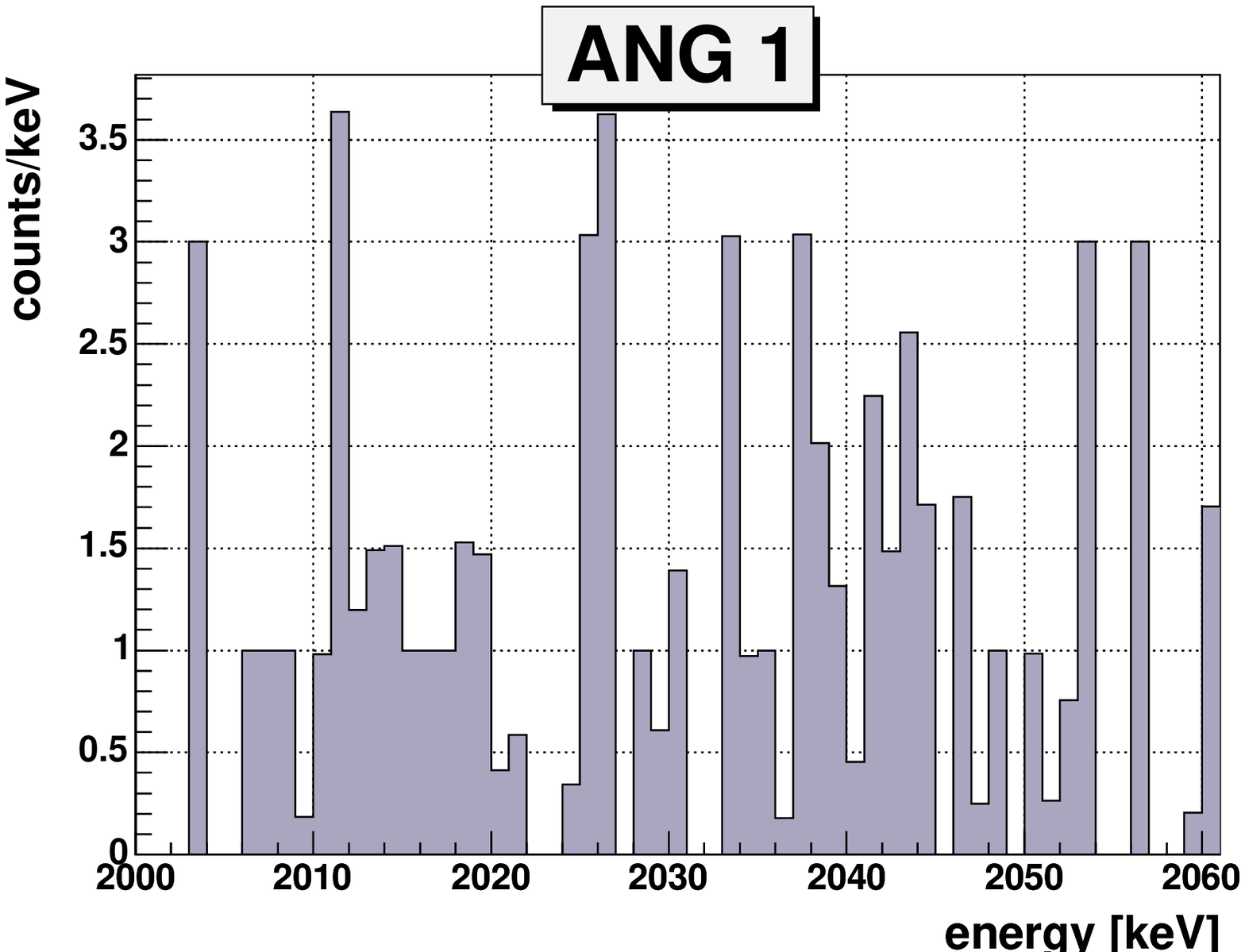}\epsfysize=40mm\epsffile{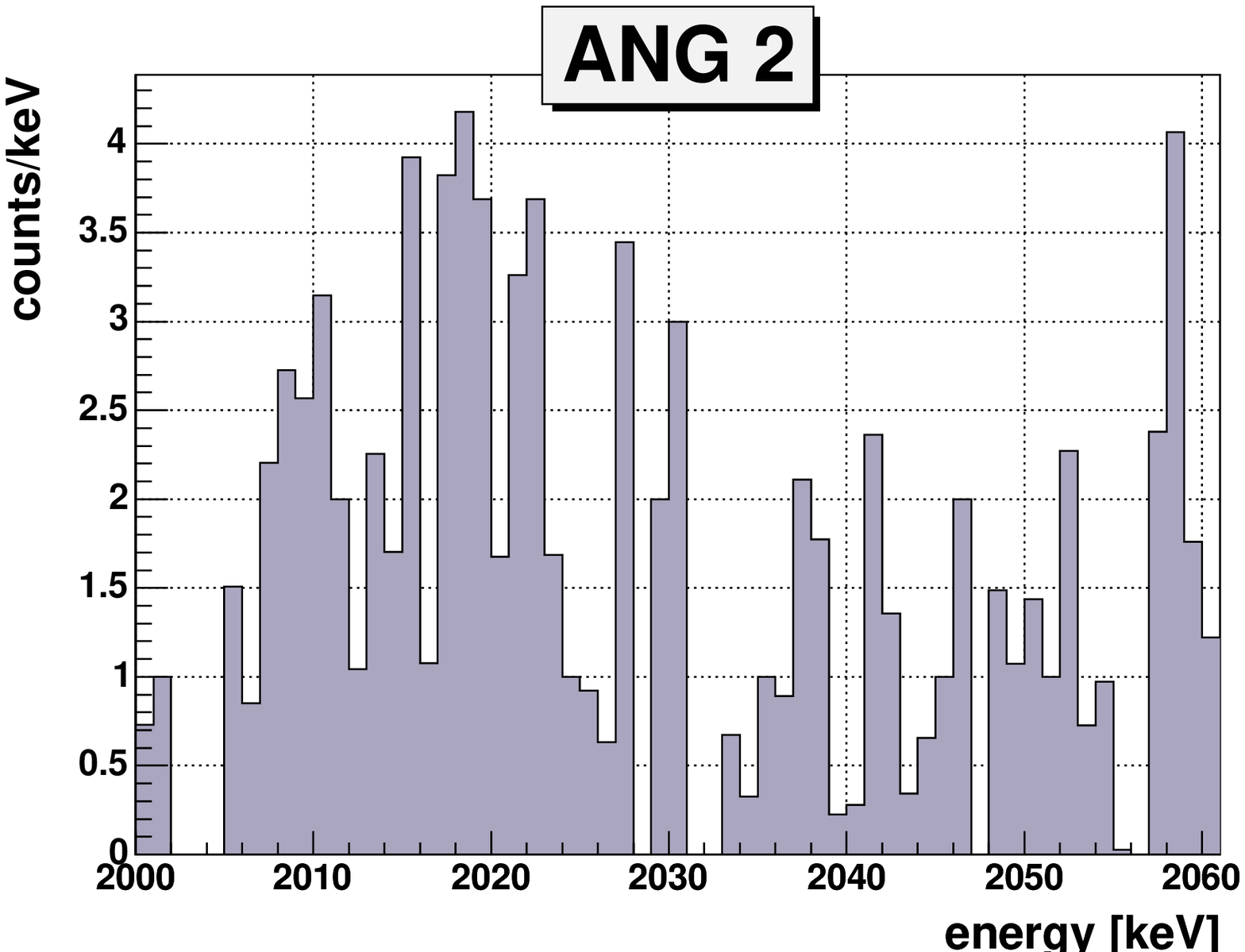}}
\epsfysize=40mm\centerline{\epsffile{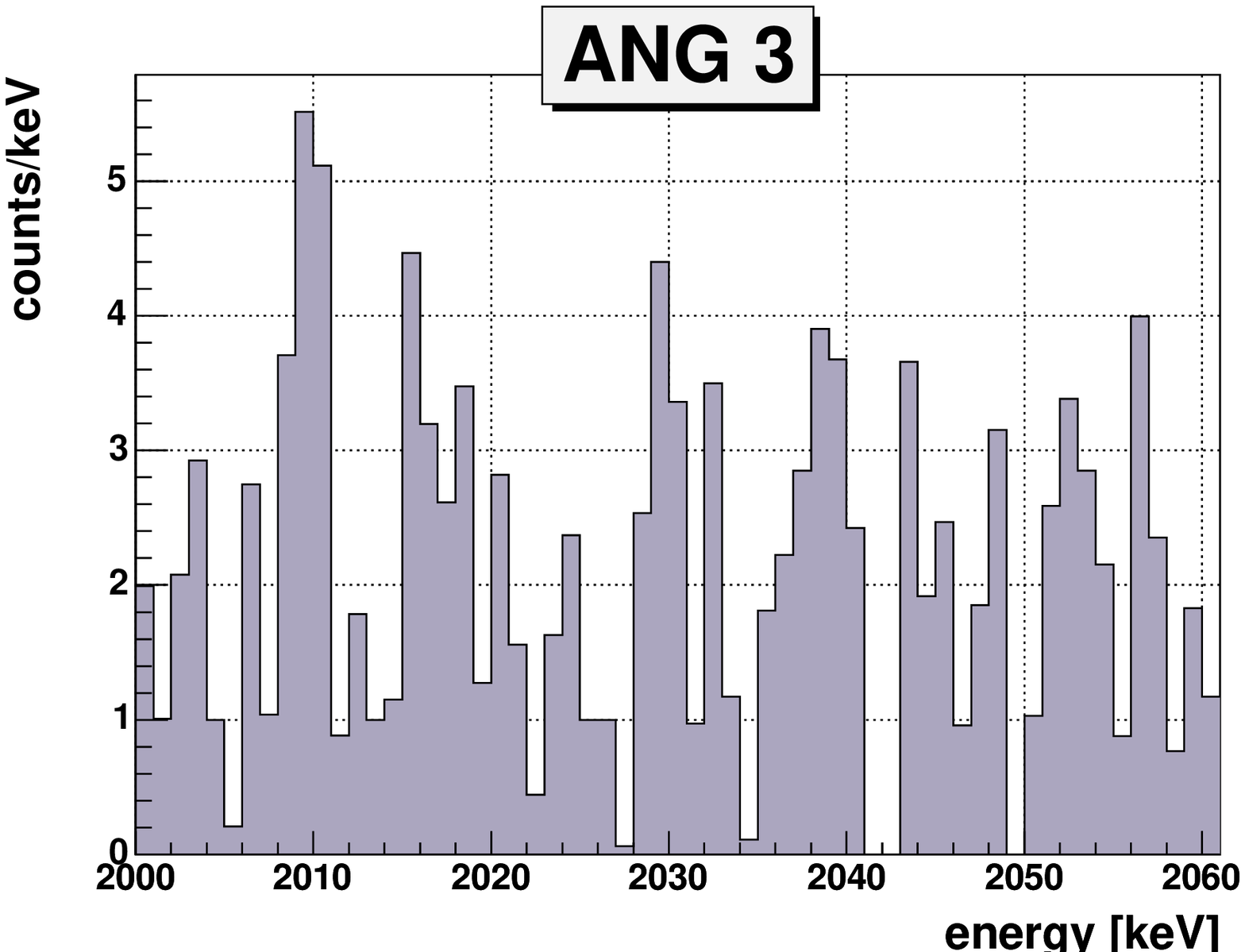}\epsfysize=40mm\epsffile{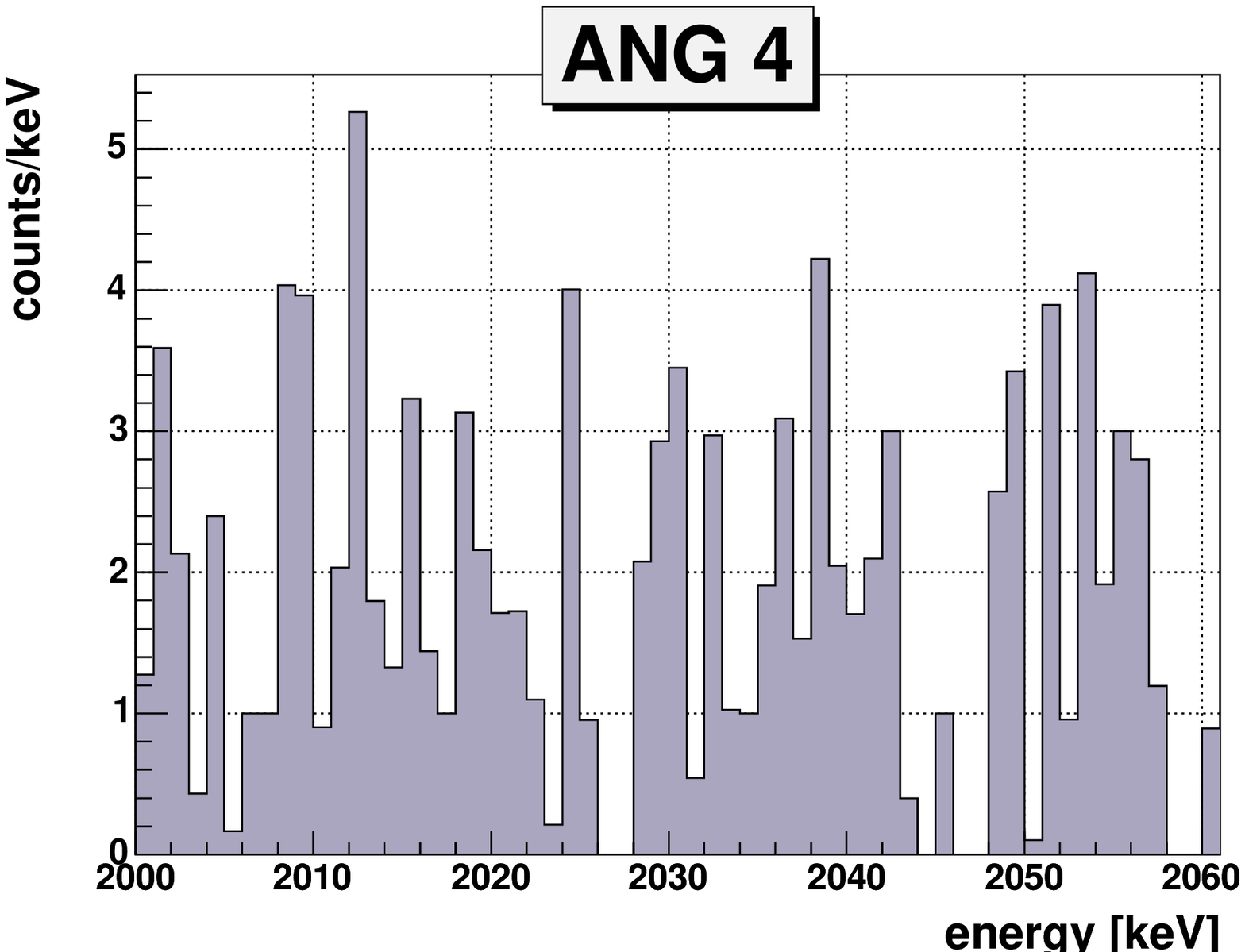}}
\epsfysize=40mm\centerline{\epsffile{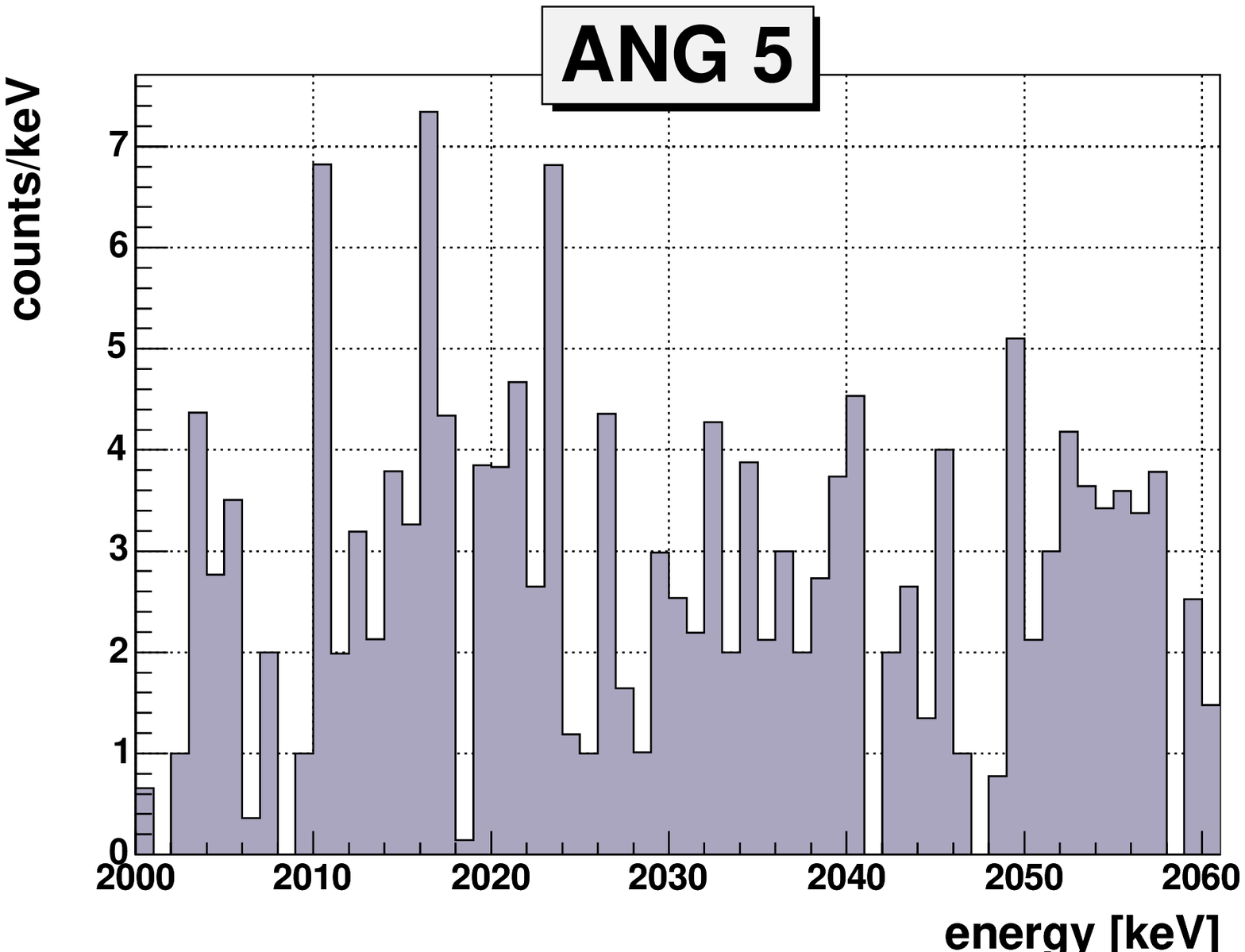}}
\caption[]{
    	The spectra of the individual five detectors 
	in the range 2000-2060\,keV for the period 
	November 1995 to May 2003. 
	The sum corresponds to the spectrum shown in Fig. 
\ref{fig:Sum90-95-03-Scan}, 
	bottom left.
}
\label{fig:IndivDet}



\vspace{0.3cm}
\epsfysize=43mm\centerline{\epsffile{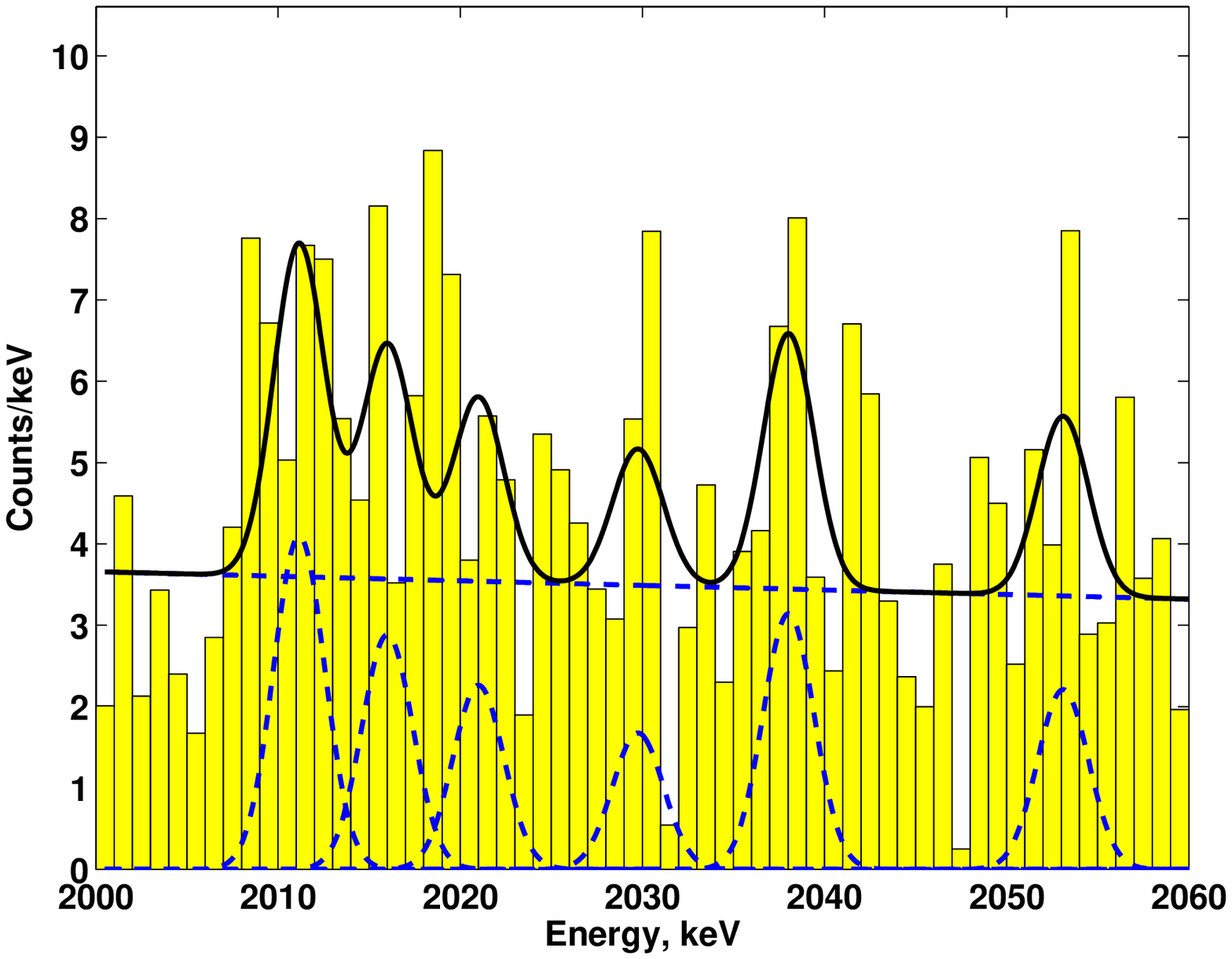}\hspace{0.5cm}\epsfysize=43mm\epsffile{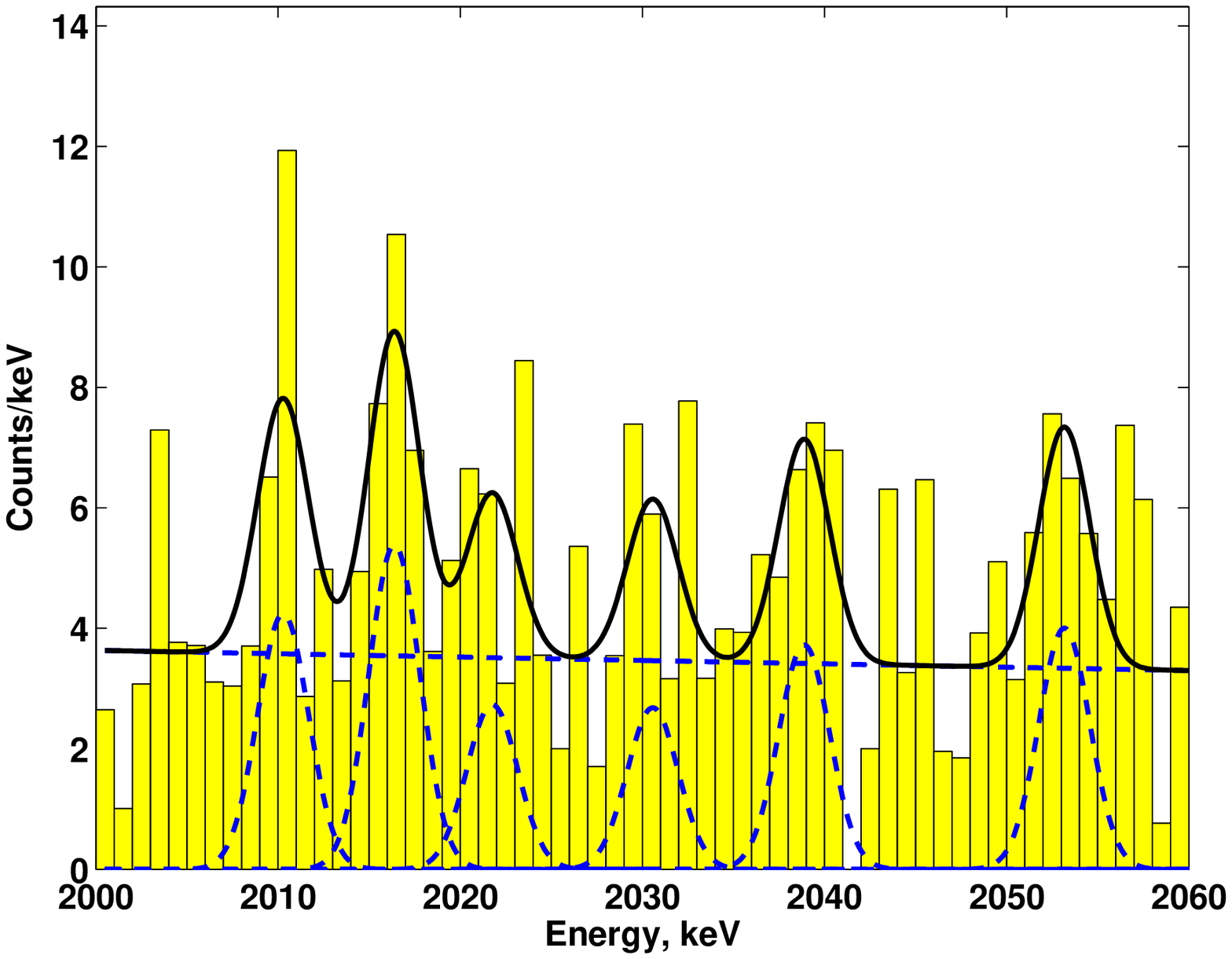}}
\caption[]{
	Spectra measured with detectors 1+2+4 (left part), 
	detectors 3+5 (right part) over the period 1995-2003. 
	The sum of the spectra is equal to that shown in Fig. 
\ref{fig:Sum90-95-03-Scan} 
	(bottom left).
}
\label{fig:124-35}
\end{figure}
\begin{figure}[ht]
\epsfysize=40mm\centerline{\epsffile{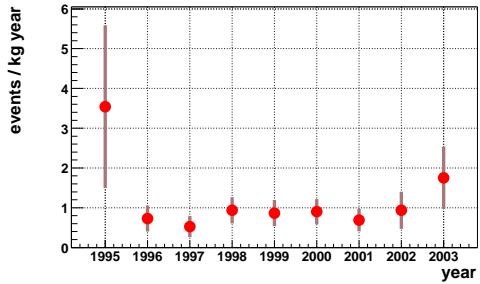}}
\caption[]{ 
	The measured {\it total} counting rates 
	in the period November 1995 to May 2003
	in the energy range 2036.5$\div$2041.0\,keV as function 
	of time (1$\sigma$ statistical errors are given).
}
\label{fig:RatesIndivDet}
\end{figure}

	In the measured spectra (Figs. 
\ref{fig:Sum90-95-03-Scan})
	we see in the range {\it around} $Q_{\beta\beta}$ the
	$^{214}{Bi}$ lines 
	at 2010.7, 2016.7, 2021.8, 2052.9\,keV,
	a line at $Q_{\beta\beta}$ and a candidate of a line 
	at $\sim$ 2030\,keV (the latter could originate, as mentioned in 
\cite{Bi-KK03-NIM,Backgr-KK03-NIM}, 
	from electron conversion of the 2118\,keV $\gamma$-line 
	from $^{114}{Bi}$. 
	On the other hand, as noted in 
\cite{Backgr-KK03-NIM}, 
	its energy differs from $Q_{\beta\beta}$ just by the K-shell 
	X-ray energies of Ge(Se) of 9.2(10.5)\,keV).
	We show the energy range 2000-2060\,keV, 
	since here a simultaneous fit can be made to safely 
	attributed lines (except the 2030\,keV line), 
	in contrast to the range we used earlier (2000-2100\,keV). 
	{\it Non-integer} numbers in the sum spectra 
	are simply a binning effect.
	While a peak-search procedure as shown in Fig. 
\ref{fig:Sum90-95-03-Scan} (bottom right)  
	was needed earlier, 
	to project out the lines from the spectrum 
\cite{KK02,KK02-Found-PN},
	they can now, with more statistics and improved analysis, 
	be directly clearly seen in the spectra.
	The large improvement of the present analysis compared 
	to our paper from 2001 
\cite{KK02},
	is clearly seen from Fig. 
\ref{fig:Sum90-95-03-Scan} 
	(top left) showing the {\it new} analysis 
	of the data 1990-2000, as performed here -- 
	to be compared to the corresponding figure in 
\cite{KK02,KK02-Found-PN}.
	One reason is, that the stricter conditions for accepting data 
	into the analysis, in particular of conditions (4) and (6) 
	in section 4, cleaned the spectrum considerably. 
	The spectrum in Fig. 
\ref{fig:Sum90-95-03-Scan}
	(top left) now corresponds to 50.57\,kg\,y 
	to be compared to 54.98\,kg\,y in 
\cite{KK02},
	for the same measuring period.
	The second reason is a better energy calibration 
	of the individual runs. The third reason 
	is the refined summing procedure of the individual data sets 
	mentioned above and the correspondingly better energy resolution 
	of the final spectrum. 
	(For more details see 
\cite{New-Anal03}). 
	The signal strength seen in the {\it individual} detectors 
	in the period 1995-2003 is shown in Fig. 
\ref{fig:IndivDet}. 
	When adding the spectra of Fig. 
\ref{fig:IndivDet} 
	for two subsets of detectors, detectors 1+2+4, 
	and 3+5,   
	both resulting spectra already clearly indicate, 
	in contradiction to a claim of 
\cite{Kurch03}, 
	the signal at Q$_{\beta\beta}$ (see Fig. 
\ref{fig:124-35} and Table 2). 
	The time distribution of the events 
	throughout the measuring time 
	and the distribution among the detectors corresponds 
	to the expectation for a constant rate, 
	and to the masses of the detectors (see Fig. 
\ref{fig:RatesIndivDet}).


	The spectra have been analyzed by {\it different methods}: 
	Least Squares Method, Maximum Likelihood Method (MLM) 
	and Feldman-Cousins Method. 
	The analysis is performed 
	{\it without subtraction of any background}. 
	We always process background-plus-signal 
	data since the difference between two Poissonian 
	variables does {\it not} produce a Poissonian distribution 
\cite{NIM99}. 
	This point is sometimes overlooked. 
	So, e.g., in 
\cite{Zdes} 
	a formula is developed making use 
	of such subtraction and as a consequence 
	the analysis given in 
\cite{Zdes} 
	provides overestimated standard errors.

	We have performed first a simultaneous fit 
	of the range 2000 - 2060\,keV 
	of the measured spectra by the nonlinear 
	least squares method, using the Levenberg-Marquardt algorithm 
\cite{Gaus-N-Meth}. 
	It is applicable in {\it any} statistics 
\cite{Zlok-88} 
	under the following conditions:
	1) relative errors asymptotic to zero, 
	2) ratio of signal to background asymptotic constant.
	It does not require 
	exact knowledge of the probability density function 
	of the data.

\clearpage


	We fitted the spectra using $n$ Gaussians 
	($n$ is equal to the number of lines, which we want to fit) 
	$G(E_i,E_j,\sigma_j)$ 
	and using different background models $B(E_i)$: 
	simulated background (linear with fixed slope) (see 
\cite{KK-Doer03}), linear, and constant. 
	The fitting function $F(E_i)$ thus is a sum 
	of Gaussians and background.

{\it Error estimate:}
	the Levenberg-Marquardt method 
\cite{Gaus-N-Meth} 
	is one of the most tested minimization algorithms, 
	finding fits most directly and efficiently. 
	It also is reasonably insensitive 
	to the starting values of the parameters. 
	This method has advantages over others of providing 
	an estimate of the full error matrix.
	The MATLAB Statistical Toolbox 
\cite{MATLAB} 
	provides functions for confidence 
	interval estimation for all parameters. 
	We tested the confidence intervals calculated 
	by the MATLAB statistical functions with numerical simulations. 
	As done earlier for other statistical methods 
\cite{KK02-Found-PN}, 
	we have simulated samples of 100 000 spectra each 
	with Poisson-distributed 
	background and a Gaussian-shaped (Poisson-distributed) 
	line of given intensity, and have investigated, 
	in how many cases we find in the analysis the known intensities 
	inside the given confidence range 
	(for details see 
\cite{New-Anal03}).
	We find the confidence levels calculated 
	by this method to be correct 
	within $\sim$ 0.3 $\sigma$.
	The second standard method, we have used to analyze 
	the measured spectra, is the Maximum Likelihood Method (MLM) 
	using the root package from CERN 
\cite{ROOT}, 
	which exploits the program MINUIT for error calculation. 
	For this purpose the program provided there 
	had to be extended for application to non-integer numbers 
	(for details see 
\cite{KK-Underthresh03}). 
	We performed also in this case a test 
	of the given confidence intervals given by this program, 
	by numerical simulations. 
	Taking into account the results, 
	the application of the MLM yields results 
	for the confidence levels consistent 
	with the Least Squares Method. 
	The third method used for analysis 
	is the Feldman-Cousins-Method (FCM) described in detail in 
\cite{RPD00-Feld}.

\section{Results}

	Fig.  
\ref{fig:Sum90-95-03-Scan} 
	shows together with the measured spectra 
	in the range around Q$_{\beta\beta}$ (2000 - 2060\,keV), 
	the fits using 
	the first of the methods described.
	In these fits a linear decreasing shape of the background as function 
	of energy was chosen, corresponding to the complete simulation 
	of the background performed in  
\cite{KK-Doer03} 
	 by GEANT4.
	In these fits the peak positions, widths and intensities 
	are determined simultaneously, and also the {\it absolute} level 
	of the background. E.g. in Fig.
\ref{fig:Sum90-95-03-Scan}(middle), 
	the {\it fitted} 
	background corresponds to (55.94 $\pm$ 3.92)\,kg\,y 
	if extrapolated from the 
	background {\it simulated} in 
\cite{KK-Doer03}
	for the measurement with 49.59\,kg\,y  
	of statistics. 
	This is almost {\it exactly} the statistical significance 
	of the present experiment (56.66\,kg y) 
	and thus a nice proof of consistency.

	For the line near Q$_{\beta\beta}$ we obtain 
	for the measuring periods 1990-2003 (1995-2003) 
	the following result: 
	energy (in keV) 2038.07$\pm$0.44 (2038.44$\pm$0.45), 
	intensity 28.75$\pm$6.86 (23.04$\pm$5.66)\,events. 
	This result is obtained assuming 
	the linearly decreasing background as described. 
	Assuming a {\it constant} background in the range 2000 - 2060\,keV 
	or keeping also the {\it slope} of a linearly 
	varying background as a free parameter,
	yields very similar results. 
	For the analysis of the other lines we refer to 
\cite{New-Anal03}.
	For the period 1990$\div$2000 (see Fig. 
\ref{fig:Sum90-95-03-Scan})  
	we obtain an intensity of 19.36$\pm$6.22\,events, 
	consistent with the 14.8\,events reported in 
\cite{KK02,KK02-Found-PN} 
	for 46.5\,kg\,y.  
	The signal at $\sim$2039\,keV
	in the full spectrum thus  
	reaches a 4.2 $\sigma$ confidence level 
	for the period 1990-2003, and of 4.1 $\sigma$ 
	for the period 1995-2003. 
	We have given a detailed comparison 
	of the spectrum measured in this experiment 
	with other Ge experiments in 
\cite{Backgr-KK03-NIM}.
	It is found that the most sensitive experiment 
	with natural Ge detectors 
\cite{Caldw91}, 
	and the first experiment using enriched (not yet high-purity) 
	$^{76}{Ge}$ detectors 
\cite{vasenko}
	find essentially the same background lines 
	($^{214}{Bi}$ etc.), but {\it no} indication 
	for the line near Q$_{\beta\beta}$.

	The fits obtained with the Maximum Likelihood Method 
	agree with the results, given above.  
	The estimations of the Feldman-Cousins Method 
	yield confidence levels 
	for the line at Q$_{\beta\beta}$ at a $\sim$ 4$\sigma$ level.
	The uncertainty in the determination of the background here  
	is larger, than in a consistent fit of the full range 
	of interest by the other two methods.

\section{Time structure of events}

	We developed for the \HM experiment some 
	{\it additional tool} of independent verification, 
	whose present status will be presented here briefly.
	The method is to exploit the time structure of the events 
	and to select $\beta\beta$ events by their pulse shape.
	Applying a method similar to that described earlier 
\cite{KK02-Found-PN}, 
	accepting only events as single site events (SSE), classified 
	by one of the neuronal net methods 
\cite{KKMaj99}, 
	and in most cases 
	simultaneously also by the second and third method 
\cite{KKMaj99} 
	mentioned in 
\cite{KK02-Found-PN}, 
	as SSE (for details see 
\cite{KK-Underthresh03}), 
	we find a 3.3$\sigma$ signal for the 
	line near Q$_{\beta\beta}$ (12.36$\pm$3.72\,events), 
	and we find the other lines, 
	at 2010.7, 2016.7 and 2021.8\,keV to be consistent 
	in intensity with {\it normal} $\gamma$-lines 
	i.e. considerably more reduced (see 
\cite{New-Anal03}). 
	When calibrating the pulse shape analysis method 
	with the 1592\,keV double escape line from $^{228}{Th}$ 
	source -- assuming that the \znbb events behave like 
	the single site events in a double escape line -- it 
	can be concluded that the line at Q$_{\beta\beta}$ 
	is consisting of SSE (for details see 
\cite{New-Anal03}).

\begin{figure}[ht]
\epsfysize=60mm\centerline{\epsffile{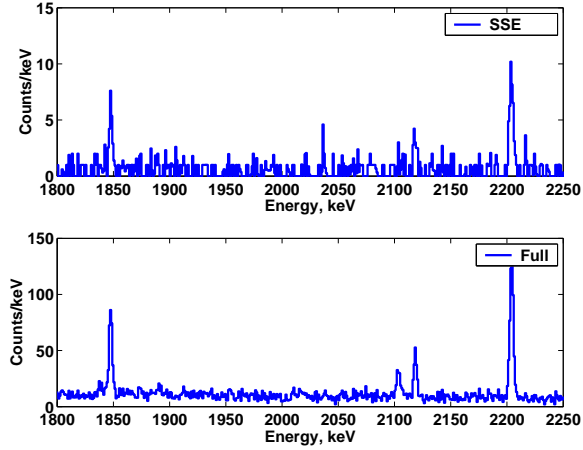}}
\caption[]{The pulse-shape selected (above) 
	and the full (below) spectrum measured 
	with detectors 2,3,4,5 from 1995-2003,  
	in the energy range 1800-2250\,keV.  
}
\label{fig:Full-NN-95-03}
\end{figure}

\begin{figure}[ht]
\epsfysize=50mm\centerline{\epsffile{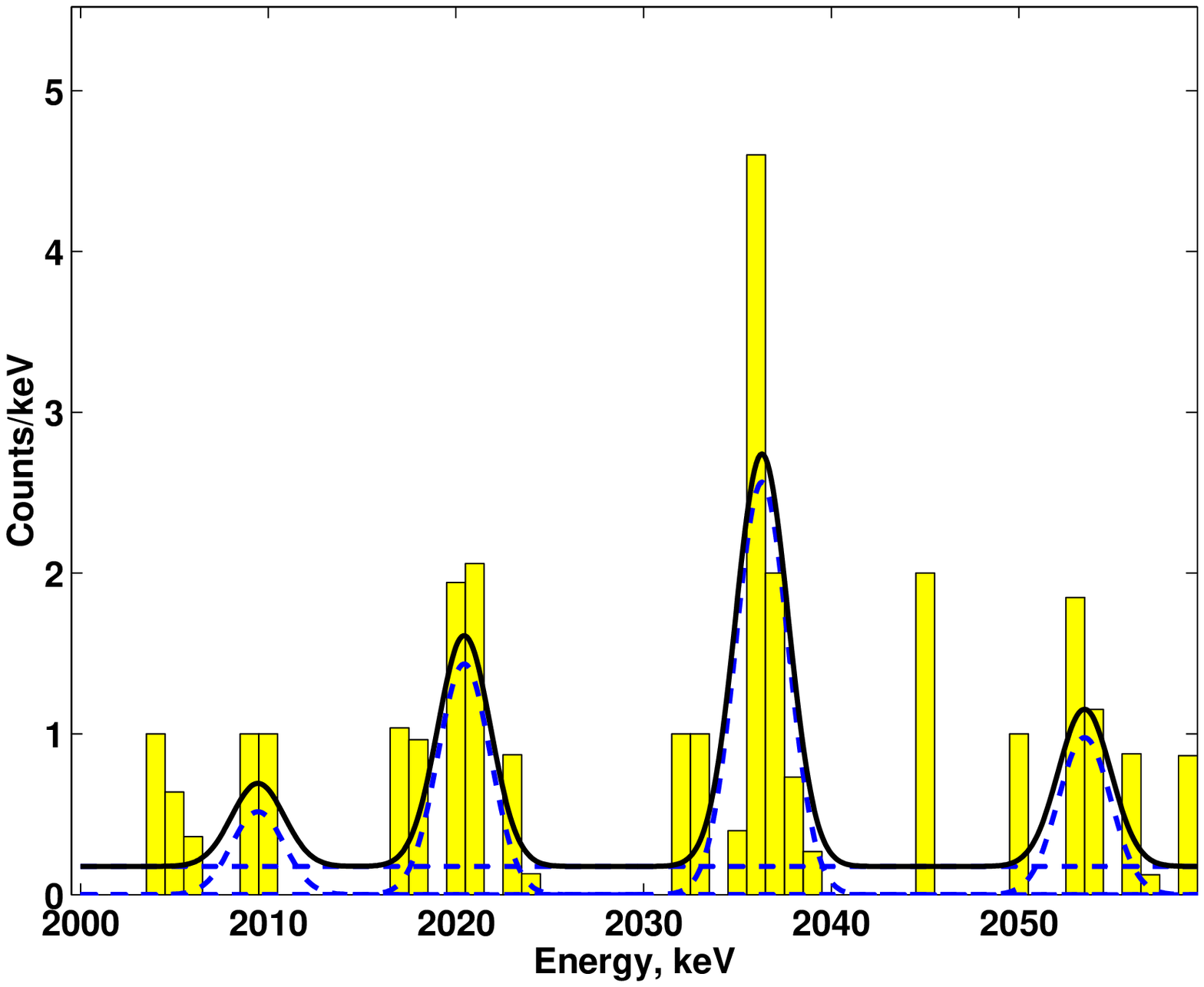}
\epsfysize=51mm\epsffile{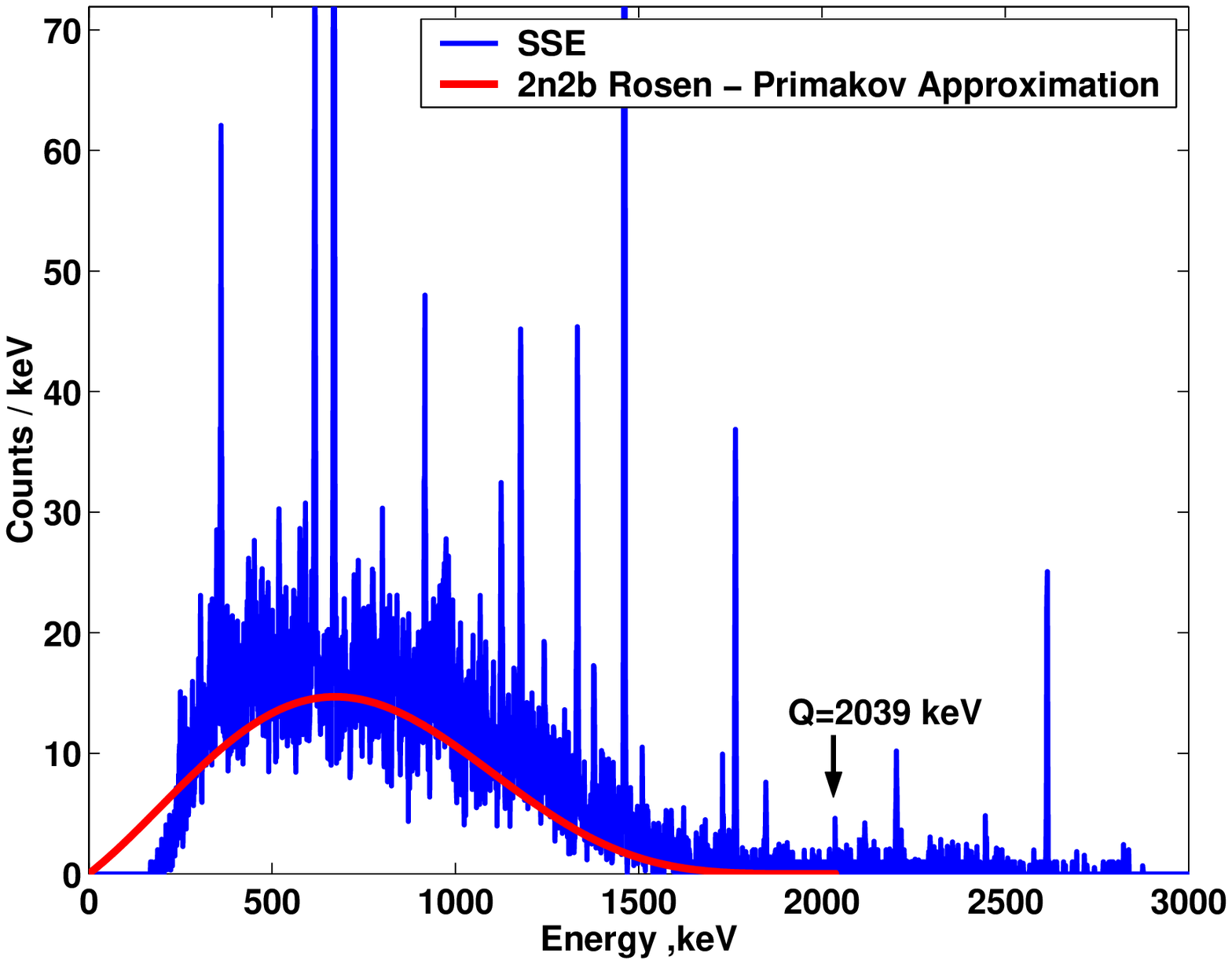}}
\caption[]{Left: 
	The pulse shape selected spectrum in the range 2000-2060\,keV, 
	taken with detectors 2,3,4,5, in the period 1995-2003  
	(compare also to Fig.
\ref{fig:Sum90-95-03-Scan} bottom, left).
	Right: the pulse-shape selected spectrum, 
	selected with the same method as in used in Fig. 
\ref{fig:Full-NN-95-03} (top) and in this figure, left part, 
	and for the same measuring period, 
	in the energy range of (100$\div$3000)\,keV.  
	The solid curve corresponds to the shape of a \tnbb spectrum. 
}
\label{fig:beta-beta}
\end{figure}

	That a neuronal net method selection 
	of the specific pulse shapes created 
	by double beta decay can be done 
	with much higher sensitivity, can be seen from Fig. 
\ref{fig:Full-NN-95-03}.
	Here a defined subclass of the shapes selected 
	above is selected, corresponding to events in part 
	of the inner volume of the detectors. 
	Except a line which sticks out sharply near Q$_{\beta\beta}$, 
	{\it all} other lines are very strongly suppressed. 
	The probability to find $\sim$7\,events 
	in two neighboring channels from background fluctuations 
	is calculated to be 0.013\%. Thus, 
	we see a line near Q$_{\beta\beta}$ at a 3.8$\sigma$ level. 
	This method also fulfills the criterium to select 
	properly the {\it continuous \tnbb spectrum} 
	(see Fig. 
\ref{fig:beta-beta}), 
	which {\it is not} done properly by the earlier 
	used PSA methods (for details see 
\cite{New-Anal03,KK-Underthresh03}). 
	We consider this as an additional proof  
	that we have observed neutrinoless double beta decay. 
	For a detailed discusion see 
\cite{KK-Underthresh03}.

	The energy of the line seen in Figs. 
\ref{fig:Full-NN-95-03},\ref{fig:beta-beta} 
	is understood to be sligtly shifted to lower energy 
	by ballistic effects (see 
\cite{KK-Underthresh03}).

	The 2039\,keV line as a single site events signal 
	cannot be the double escape line of a $\gamma$-line 
	whose full energy peak would be expected at 3061\,keV 
	where no indication of a line is found in the spectrum 
	measured up to 8\,MeV (see 
\cite{New-Anal03}). 
	Summarizing, 
	the investigation of $\beta\beta$-like events 
	is proving    
	a line near Q$_{\beta\beta}$ consisting 
	of \znbb events.


\vspace{0.3cm}
\begin{table}[h]
\caption[]{\label{Results}
	Half-life for the neutrinoless decay mode 
	and the deduced effective neutrino mass 
	from the HEIDELBERG-MOSCOW experiment 
	(the nuclear matrix element of 
\cite{Sta90} 
	is used, see text). 
	Shown are in addition to various accumulated total 
	measuring times also the results for four 
	{\it non-overlapping} data sets: 
	the time periods 11.1995-09.1999 and 09.1999$\div$05.2003 for 
	{\it all} detectors, and the time period 1995$\div$2003 
	for two sets of detectors: 1+2+4, and 3+5.
	*) denotes best value.
}

\vspace{0.3cm}
\begin{center}
\newcommand{\m}{\hphantom{$-$}}
\renewcommand{\arraystretch}{1.1}
\setlength\tabcolsep{5.9pt}
\begin{tabular}{|c|c|c|c|c|}
\hline
Significan-	&	Detectors		
&	${\rm T}_{1/2}^{0\nu}	~~[y]$	
& $\langle m \rangle $ [eV]	&	Conf. \\
	ce $[ kg\,y ]$	&	&	(3$\sigma$ range) 
&		(3$\sigma$ range)& level ($\sigma$)\\
\hline
\multicolumn{5}{|c|}{\it Period 1990 $\div$ 2003}\\
\hline
	71.7	
&	1,2,3,4,5	
&	$(0.69 - 4.18) \times 10^{25}$	
& (0.24 - 0.58)	
& 	4.2	\\
 		&	
&	$1.19 \times 10^{25}$$^*$	
& 0.44$^*$ 	
& 	\\
\hline
\multicolumn{5}{|c|}{\it Period 1995 $\div$ 2003}\\
\hline
 	56.66	
&	1,2,3,4,5
&	$(0.67 - 4.45) \times 10^{25}$	
& (0.23 - 0.59)
& 4.1	\\
 &	&	$1.17 \times 10^{25}$$^*$
& 0.45$^*$ 
& 		\\
\hline
 	51.39	
&	2,3,4,5
&	$(0.68 - 7.3) \times 10^{25}$	
& (0.18 - 0.58)
& 3.6	\\
 &	&	$1.25 \times 10^{25}$$^*$	
& 0.43$^*$ 
& 		\\
\hline
 	42.69	
&	2,3,5
&	$(0.88 - 4.84) \times 10^{25}$	
& (0.22 - 0.51)
& 2.9	\\
 	&	
&	(2$\sigma$ range)	& (2$\sigma$ range)	& 	\\
 &	&	$1.5 \times 10^{25}$$^*$	
& 0.39$^*$ 
& 		\\
\hline
 	51.39 (SSE)	
&	2,3,4,5
&	$(1.04 - 20.38) \times 10^{25}$	
& (0.11 - 0.47)
& 3.3	\\
 &	&	$1.98 \times 10^{25}$$^*$ 	
& 0.34$^*$  
& 		\\
\hline
 	28.21 	
&	1,2,4
&	$(0.67 - 6.56) \times 10^{25}$	
& (0.19 - 0.59)
& 2.5	\\
 	&	
&	(2$\sigma$ range)	& (2$\sigma$ range)	& 	\\
 &	&	$1.22 \times 10^{25}$$^*$ 	
& 0.44$^*$  
& 		\\
\hline
 	28.35	
&	3,5
&	$(0.59 - 4.29) \times 10^{25}$	
& (0.23 - 0.63)
& 2.6	\\
 	&	
&	(2$\sigma$ range)	& (2$\sigma$ range)	& 	\\
 &	&	$1.03 \times 10^{25}$$^*$ 	
& 0.48$^*$  
& 		\\
\hline
\multicolumn{5}{|c|}{\it Period 1990 $\div$ 2000}\\
\hline
 	50.57	
&	1,2,3,4,5
&	$(0.63 - 3.48) \times 10^{25}$	
& (0.08 - 0.61)
& 3.1	\\
 &	&	$1.24 \times 10^{25}$$^*$
& 0.43$^*$ 
& 		\\
\hline
\multicolumn{5}{|c|}{\it Period 1995 $\div$ 09.1999}\\
\hline
 	26.59	
&	1,2,3,4,5
&	$(0.43 - 12.28) \times 10^{25}$	
& (0.14 - 0.73)
& 3.2	\\
 &	&	$0.84 \times 10^{25}$$^*$
& 0.53$^*$ 
& 		\\
\hline
\multicolumn{5}{|c|}{\it Period 09.1999 $\div$ 05.2003}\\
\hline
 	30.0	
&	1,2,3,4,5
&	$(0.60 - 8.4) \times 10^{25}$	
& (0.17 - 0.63)
& 3.5	\\
 &	&	$1.12 \times 10^{25}$$^*$
& 0.46$^*$ 
& 		\\
\hline
\end{tabular}
\end{center}
\end{table}


\clearpage
\section{Half-Life of Neutrinoless Double Beta Decay of $^{76}{Ge}$}

	On the basis of the findings presented above 
	we translate the observed number 
	of events into the half-life for neutrinoless double beta decay. 
	In Table 
\ref{Results}
	we give the half-lives deduced from 
	the full data sets taken in 1995-2003 
	and in 1990-2003\,keV 
	and of some partial data sets, including also the values deduced 
	from the single site spectrum. 
	In the latter case we do conservatively not apply, 
	a calibration factor. 
	Also given are the effective neutrino masses.
	The result obtained is consistent 
	with the results we reported in 
\cite{KK02,KK02-Found-PN}, 
	and also with the limits given earlier in 
\cite{HDM01}.  
	Concluding we confirm, 
	with 4.2$\sigma$ (99.9973$\%$ c.l.) probability, 
	our claim from 2001 
\cite{KK02,KK02-Found-PN} 
	of first evidence 
	for the neutrinoless double beta decay mode.

\section{Consequences}
%

	Consequence of the above result is, that 
	lepton number is not conserved. 
	Further the neutrino 
	is a Majorana particle see, e.g. 
\cite{Sch82a}.
	Both of these conclusions are {\it independent of any} 
	discussion of nuclear matrix elements.
	The matrix element enters when we derive 
	a {\it value} for the effective neutrino mass 
	- making the {\it most natural assumption} 
	that the \znbb decay amplitude 
	is dominated by exchange of a massive Majorana neutrino. 
	The half-life for the neutrinoless decay mode 
	is under this assumption given by 
\cite{Sta90}

$[T^{0\nu}_{1/2}(0^+_i \rightarrow 0^+_f)]^{-1}= C_{mm} 
\frac{\langle m \rangle^2}{m_{e}^2}
+C_{\eta\eta} \langle \eta \rangle^2 + C_{\lambda\lambda} 
\langle \lambda \rangle^2 +C_{m\eta} \langle \eta \rangle \frac{\langle m \rangle}{m_e} 
%
+ C_{m\lambda}
\langle \lambda \rangle \frac{\langle m_{\nu} \rangle}{m_e}+C_{\eta\lambda}
\langle \eta \rangle \langle \lambda \rangle,$
\hspace{1.cm}$\langle m \rangle = 
|m^{(1)}_{ee}| + e^{i\phi_{2}} |m_{ee}^{(2)}|
+  e^{i\phi_{3}} |m_{ee}^{(3)}|~$,

	where $m_{ee}^{(i)}\equiv |m_{ee}^{(i)}| \exp{(i \phi_i)}$ 
	($i = 1, 2, 3$)  are  the contributions 
	to the effective mass $\langle m \rangle$
	from individual mass eigenstates, 
	with  $\phi_i$ denoting relative Majorana phases connected 
	with CP violation, and
	$C_{mm},C_{\eta\eta}, ...$ denote nuclear matrix elements squared. 
	Ignoring contributions from right-handed weak currents on the 
	right-hand side of the above equation, only the first term remains.

	Using the nuclear matrix element from 
\cite{Sta90}, 
	we conclude from the measured half-life  
	given in Table 2 the effective mass 
	$\langle m \rangle $ 
	to be $\langle m \rangle $ 
	= (0.2 $\div$ 0.6)\,eV (99.73$\%$ c.l.).
	The matrix element given by 
\cite{Sta90}
	was the {\it prediction closest to} the {\it later} measured \tnbb
	decay half-life 
	of $({1.74} ^{+0.18}_{-0.16})\times 10^{21}$\,y 
\cite{KK-Doer03,HDM97}. 
	It underestimates the 2$\nu$ matrix elements by 32\% 
	and 
	thus these calculations will also underestimate (to a smaller extent) 
	the matrix element for \znbb decay, 
	and consequently correspondingly overestimate 
	the (effective) neutrino mass. 
	Allowing conservatively for an uncertainty of the nuclear 
	matrix element of $\pm$ 50$\%$, 
	the range for the effective mass may widen 
	to $\langle m \rangle $ = (0.1 - 0.9)\,eV (99.73\% c.l.).

\clearpage
	With the value deduced for the effective neutrino mass,  
	the HEIDELBERG-MOSCOW experiment excludes several 
	of the neutrino mass scenarios 
	allowed from present neutrino oscillation experiments
	(see Fig. 1, in  
\cite{KK-Sark-WMAP03-Drugie})
	- allowing only for degenerate mass scenarios 
\cite{KK-Beyond03-BB}.

	Assuming other mechanisms to dominate the \znbb~ decay amplitude, 
	which have been studied extensively in recent years,  
	the result allows to set stringent limits on parameters of SUSY 
	models, leptoquarks, compositeness, masses of heavy neutrinos, 
	of the right-handed W boson, and possible violation of Lorentz 
	invariance and equivalence principle in the neutrino sector. 
	For a discussion and for references we refer to 
\cite{KK60Y,KKS-INSA02}.

\section{Conclusion - Perspectives}
	Concluding, 
	evidence for a signal at Q$_{\beta\beta}$ on a confidence 
	level of 4.2$\sigma$ has been observed confirming our earlier claim. 
	On the basis of our pulse shape analysis this 
	can be interpreted as evidence for neutrinoless 
	double beta decay of $^{76}{Ge}$, and thus 
	for {\it total} lepton number nonconservation, 
	and for a non-vanishing Majorana neutrino mass.
	Recent information from many {\it independent} sides seems 
	to condense now to a nonvanishing neutrino mass of the order 
	of the value found by the \HM experiment.
	This is the case for the results from 
	CMB, LSS, neutrino oscillations, particle theory and cosmology 
	(for a detailed discussion see 
\cite{New-Anal03}). 
	To mention a few examples: 
	Neutrino oscillations require in the case 
	of degenerate neutrinos common mass eigenvalues 
	of m $>$ 0.04\,eV.
	An analysis of CMB, large scale structure and X-ray 
	from clusters of galaxies yields a 'preferred' value 
	for $\sum m_\nu$ of 0.6\,eV 
\cite{Allen03-Wmap}. 
	WMAP yields $\sum m_\nu$ $<$ 1.0\,eV 
\cite{Hannes03},  
	SDSS yields $\sum m_\nu$ $<$ 1.7\,eV 
\cite{Barger03}.  
	Theoretical papers require degenerate neutrinos 
	with m $>$ 0.1 and 0.2\,eV 
\cite{BMV02,Moh03}, 
	and 
	the recent alternative cosmological concordance 
	model requires relic neutrinos with mass of order of eV 
\cite{S-Sarkar03}. 
	The Z-burst scenario for ultra-high energy cosmic 
	rays requires $m_\nu$$\sim$ 0.4\,eV 
\cite{Farj00-04keV-Fodor}, 
	and also non-standard model (g-2) has been connected 
	with degenerate neutrino masses $>$0.2\,eV 
\cite{MaRaid01}.
	It has been discussed that the Majorana nature of the neutrino 
	may tell us that spacetime does realize a construct 
	that is central to construction of supersymmetric theories 
\cite{Ahl96}.

	Looking into the future, there is some hope from 
	the numbers given in Table 2, that the future tritium 
	experiment KATRIN 
\cite{Trit03}
	may see a positive signal, if tritium decay can solve 
	this problem at all (see 
\cite{Kirch04}). 
	
	From future double beta projects to improve 
	the present accuracy of the effective neutrino mass  
	one has to require that they should be able 
	to differentiate between a $\beta$ and a $\gamma$ signal, or 
	that the tracks of the emitted electrons should be measured. 
	At the same time, as is visible from the present information,  
	the energy resolution should be {\it at least} 
	in the order of that of Ge semiconductor detectors.  
	If one wants 
	to get {\it independent additional} insight into the neutrinoless 
	double beta  decay process, 
	one would probably wish to see the effect in {\it another} 
	isotope, which would then simultaneously give additional 
	information also on the nuclear matrix elements. 
	In view of the above remarks, future efforts to obtain 
	{\it deeper} information on the process 
	of neutrinoless double beta decay, would require   
	{\it a new experimental approach, different from all, 
	what is at present persued}.

\section{Acknowledgement}

%
	The authors would like to thank all colleagues, 
	who have contributed to the experiment over the last 15\,years. 
	They are particularly grateful to C. Tomei 
	for her help in the early calibration of the pulse shape methods 
	with the Th source and to Mr. H. Strecker 
	for his invaluable technical support. 
	Our thanks extend also to the technical staff of the 
	Max-Planck Institut f\"ur Kernphysik and 
	of the Gran Sasso Underground Laboratory.  
	We acknowledge the invaluable support from BMBF 
	and DFG, and LNGS of this project.
	We are grateful to the former State Committee of Atomic 
	Energy of the USSR for providing the enriched material 
	used in this experiment.


\end{document}